\pdfoutput=1
\documentclass[twocolumn,
superscriptaddress,aps,
prd,
preprintnumbers,amsmath,amssymb,nofootinbib,
]{revtex4-1}

\usepackage{placeins}
\usepackage{amsmath}
\usepackage{amssymb}
\usepackage{amsfonts}
\usepackage{dsfont}
\usepackage{graphicx}
\usepackage[usenames,dvipsnames,svgnames,table]{xcolor}
\usepackage{xfrac}
\usepackage{comment}
\usepackage{pifont}
\usepackage{physics}
\usepackage{hyperref}
\usepackage{bm}
\usepackage{enumitem}
\usepackage[normalem]{ulem}

\definecolor{rossoferrari}{HTML}{D9073D}
\definecolor{mediumblue}{HTML}{0000CD}
\definecolor{forestgreen}{HTML}{228B22}
\definecolor{desy_blue}{HTML}{009EE2}
\definecolor{desy_orange}{HTML}{FD8800}
\definecolor{peera_green}{HTML}{008B8B}
\definecolor{peera_orange}{HTML}{B22222}
\definecolor{light_pink}{rgb}{1,0.4,0.4}
\definecolor{light_blue}{rgb}{0.284602,0.317763,0.963947}
\definecolor{peera_col}{RGB}{240, 94, 28}
\definecolor{blue_col}{RGB}{0,92,175}
\definecolor{red_col}{RGB}{203,64,66}
\usepackage{hyperref}
\hypersetup{linktocpage,
    colorlinks=true,
    linkcolor=red_col,
    filecolor=Mahogany,      
    urlcolor=blue_col,
    citecolor=blue_col,
}

\begin{document}


\title{
Ultrahigh frequency primordial gravitational waves beyond the kHz: \\The case of cosmic strings 
}

\author{G{\'e}raldine Servant} 
\email{geraldine.servant@desy.de}
\affiliation{Deutsches Elektronen-Synchrotron DESY, Notkestra{\ss}e 85, 22607 Hamburg, Germany}
\affiliation{II. Institute of Theoretical Physics, Universit\"{a}t  Hamburg, 22761, Hamburg, Germany}
\affiliation{Theoretical Physics Department, CERN, 1211 Geneva 23, Switzerland}

\author{Peera Simakachorn}
\email{peera.simakachorn@ific.uv.es}
\affiliation{Instituto de F\'isica Corpuscular (IFIC), Universitat de Val\`{e}ncia-CSIC,\\C/ Catedrático José Beltrán 2, E-46980, Paterna, Spain}

\preprint{DESY-23-202}
\preprint{CERN-TH-2023-226}

\date{May 23, 2024}


\begin{abstract}
\noindent
We investigate gravitational-wave backgrounds (GWBs) of primordial origin that would manifest only at ultra-high frequencies, from kilohertz to 100 gigahertz, and leave no signal at either LIGO, Einstein Telescope, Cosmic Explorer, LISA, or pulsar-timing arrays. We focus on GWBs produced by cosmic strings and make predictions for the GW spectra scanning over high-energy scale (beyond $10^{10}$ GeV) particle physics parameters.
Signals from local string networks can easily be as large as the Big Bang nucleosynthesis/cosmic microwave background bounds, with a characteristic strain as high as $10^{-26}$ in the 10 kHz band, offering prospects to probe grand unification physics in the $10^{14}-10^{17}$ GeV energy range. In comparison, GWB from axionic strings is suppressed (with maximal characteristic strain $\sim 10^{-31}$) due to the early matter era induced by the associated heavy axions.
 We estimate the needed reach of hypothetical futuristic GW detectors to probe such GWB and, therefore, the corresponding high-energy physics processes. Beyond the information of the symmetry-breaking scale, the high-frequency spectrum encodes the microscopic structure of the strings through the position of the UV cutoffs associated with cusps and kinks, as well as potential information about friction forces on the string. The IR slope, on the other hand, reflects the physics responsible for the decay of the string network. We discuss possible strategies for reconstructing the scalar potential, particularly the scalar self-coupling, from the measurement of the UV cutoff of the GW spectrum.

\end{abstract}


\maketitle


{\small \tableofcontents}


\section{Primordial GWB at ultra-high frequencies}

The landscape of gravitational waves (GW) in the ultra-high frequency (UHF) regime, above the kHz, is beyond the sensitivities of the present terrestrial experiments:  LIGO/Virgo/KAGRA \cite{LIGOScientific:2014qfs,LIGOScientific:2019vic}, and the planned: Einstein Telescope (ET) \cite{Hild:2010id, Punturo:2010zz} and Cosmic Explorer (CE) \cite{LIGOScientific:2016wof}. It is compelling because it is clean from the standard astrophysical GW signals, such as binaries of massive objects \cite{Rosado:2011kv,Sesana:2016ljz,Lamberts:2019nyk,Robson:2018ifk,KAGRA:2021kbb,KAGRA:2021vkt,Babak:2023lro} and would, in principle, be the reserved domain of early-Universe signals, ranging from primordial inflation \cite{Grishchuk:1974ny,Starobinsky:1979ty,Rubakov:1982df,Bartolo:2016ami}, thermal plasma \cite{Ghiglieri:2015nfa, Ghiglieri:2020mhm, Ringwald:2020ist, Ghiglieri:2022rfp, Ghiglieri:2024ghm}, first-order phase transitions \cite{Caprini:2015zlo,Caprini:2019egz,Hindmarsh:2020hop,Gouttenoire:2022gwi,Athron:2023xlk}, 
topological defects \cite{Vilenkin:2000jqa, Blanco-Pillado:2017oxo, Auclair:2019wcv, Gouttenoire:2019kij}, 
primordial black-holes \cite{Anantua:2008am,Dolgov:2011cq,Dong:2015yjs,Franciolini:2022htd,Gehrman:2022imk,Gehrman:2023esa}
and preheating \cite{Easther:2006gt,Garcia-Bellido:2007nns,Garcia-Bellido:2007fiu,Dufaux:2007pt,Figueroa:2022iho,Barman:2023ymn}; see reviews \cite{ Caprini:2018mtu,Renzini:2022alw,LISACosmologyWorkingGroup:2022jok,Simakachorn:2022yjy}.
The recently launched ``UHF-GW Initiative'' reviewed the detector concepts that have been proposed to explore this almost uncharted territory in Ref.~\cite{Aggarwal:2020olq}.

\begin{figure*}[t]
	\centering
 \includegraphics[width=\linewidth]{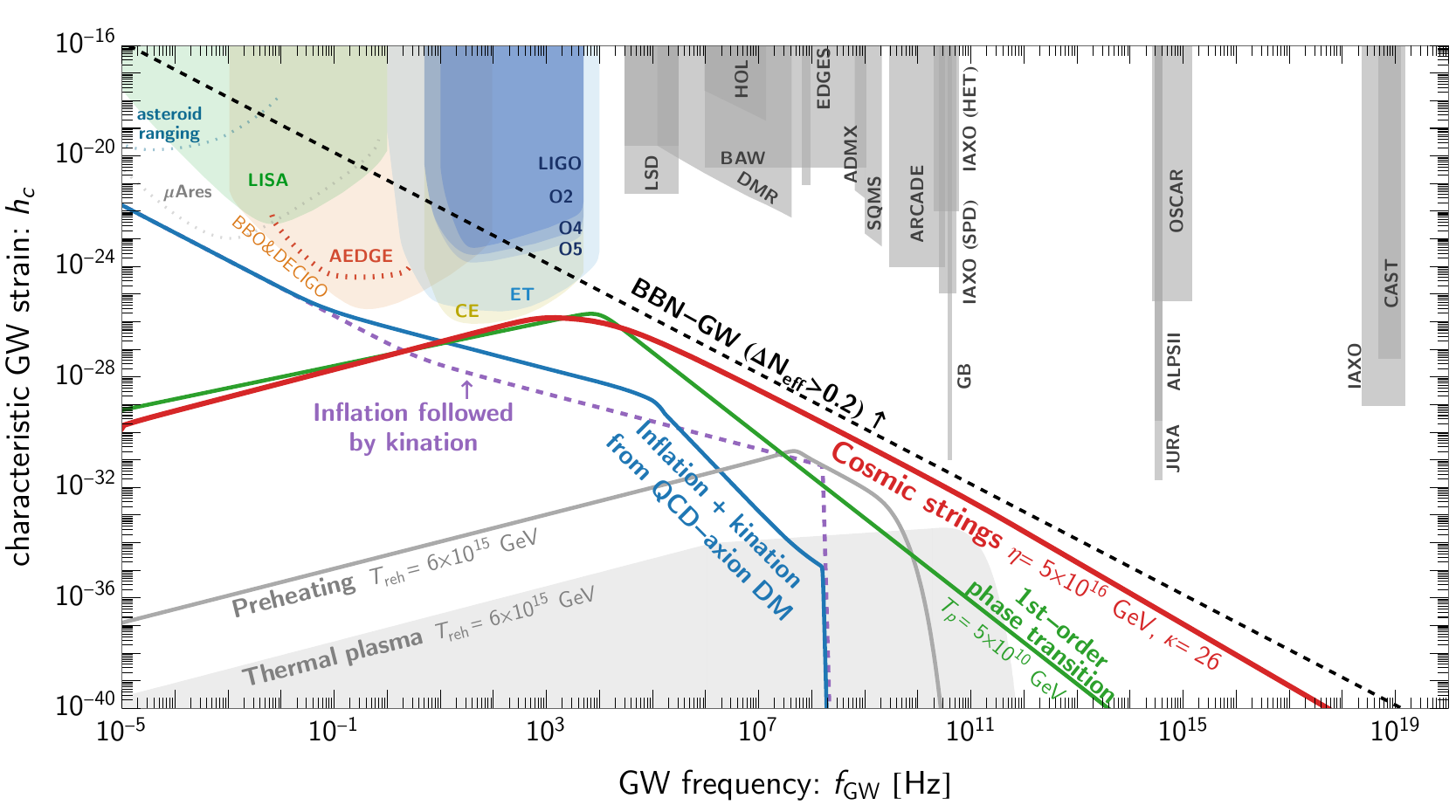}\\[-1em]
\caption{Benchmark GWBs of primordial origin  with large amplitude above kHz frequencies, compared to sensitivities of existing and planned experiments below the kHz
\cite{LISA:2017pwj, LISACosmologyWorkingGroup:2022jok, Yagi:2011wg, AEDGE:2019nxb, KAGRA:2013rdx, LIGOScientific:2014qfs, LIGOScientific:2019vic, Hild:2010id, Punturo:2010zz, LIGOScientific:2016wof,Blas:2021mqw,Fedderke:2021kuy}
as well as experiments sensitive at frequencies above the kHz from \cite{Aggarwal:2020olq} (in shaded gray). The green line is associated with a very strong first-order phase transition \cite{Caprini:2015zlo} ($\beta/H=7$, $\alpha=10$)  at a temperature $T\sim 10^{10}$ GeV (compatible with a Peccei-Quinn phase transition with axion decay constant $f_a\sim 10^{10}$ GeV for instance \cite{VonHarling:2019rgb}). Interestingly, the irreducible background from inflation with inflationary scale $E_{\rm inf} \simeq 10^{16}$ GeV can be amplified if inflation is followed by kination (purple line) \cite{Gouttenoire:2021jhk} or if a kination era is induced much later by the rotating QCD axion DM field (blue line) \cite{Gouttenoire:2021wzu,Gouttenoire:2021jhk, Co:2021lkc}. Local cosmic strings can generate a signal (in red) as large as the BBN bound \eqref{eq:Neff_bound}, that also uniquely goes beyond $10^9$ Hz. The gray line shows the signal from preheating \cite{Figueroa:2022iho} corresponding to an inflaton mass $M\simeq M_{\rm pl}$ with a coupling $g = 10^{-3}$ to the thermal bath.  Similar but suppressed GW spectra can come from the fragmentation of a scalar field, which is not the inflaton \cite{Kitajima:2018zco, Chatrchyan:2020pzh, Eroncel:2022vjg}. The lower gray shaded region is the spectrum from the Standard thermal plasma \cite{Ghiglieri:2015nfa, Ghiglieri:2020mhm, Ringwald:2020ist}, assuming a reheating temperature $T_{\rm reh}\simeq 6 \times 10^{15}$ GeV.}
    \label{fig:money_plot_intro}
\end{figure*}

The above primordial signals contribute to a stochastic gravitational-wave background (GWB) characterized by its frequency power spectrum, commonly expressed as the GW fraction of the total energy density of the Universe today $\Omega_{\rm GW}h^2$.
It can be related to the characteristic strain $h_c$ of GW by \cite{Caprini:2018mtu}
\begin{align}
    h_c \simeq 1.26 \times 10^{-18} ({\rm Hz}/f_{\rm GW})\sqrt{\Omega_{\rm GW}h^2}. \end{align}
Its characteristic frequency is related to the moment when GW was emitted, and
its amplitude is typically small\footnote{Except the signals resulting from a modified equation of state of the Universe such as kination or stiff eras \cite{Gouttenoire:2021jhk} or extremely strong first-order phase transitions.} ($\Omega_{\rm GW}h^2 \lesssim \Omega_{r}h^2 \simeq 4 \cdot 10^{-5}$ \cite{Planck:2018vyg}, where $\Omega_{r}$ is the fraction of energy density in radiation).

The frequency range of cosmological GWB is linked to the size of the source, which is limited to the horizon size by causality. The frequency today of a GW produced with wavelength $\lambda_{\rm GW} \leq H^{-1}(T)$  when the Universe had temperature $T$ (assuming radiation domination for the GW considered in this paper) is
\begin{align}
    f_{\rm GW} \simeq 1 ~ {\rm kHz} \left[\frac{H^{-1}(T)}{\lambda_{\rm GW}}\right]\left(\frac{T}{{10^{10} ~ \rm GeV}}\right),
    \label{eq:GW_freq_horizon_temperature}
\end{align}
where $H$ is the Hubble expansion rate, and $f_{\rm GW} = \lambda_{\rm GW}^{-1}[a(T)/a_0]$ with $a$ being the scale factor of cosmic expansion.
For instance, the irreducible GWs produced during inflation that re-enter the horizon at temperature $T$ have $\lambda_{\rm GW} \sim H^{-1}$.
On the other hand, GWs from first-order phase transitions have $\lambda_{\rm GW} $ that is roughly the bubbles' size, typically of the order $\mathcal{O}(10^{-3}-10^{-1}) H^{-1}$. GWs produced from the thermal plasma are produced maximally at $\lambda_{\rm GW} \sim T^{-1}$, such that the signal generated at any $T$ is peaked at $f_{\rm GW} \sim \mathcal{O}(10)$ GHz. Finally, for cosmic strings, $\lambda_{\rm GW}$ relates to the string-loop size, which is fixed by the Hubble size; see Eq.~\eqref{eq:freq_temp_relation} for the precise relation. Therefore, apart from the thermal plasma source, the highest GW frequencies are associated with the earliest moments in our Universe's history, and the maximum reheating temperature of the Universe $T_{\rm reh} \leq T_{\rm reh}^{\rm max} \simeq 6 \times 10^{15}$ GeV \cite{Planck:2018jri} bounds  $f_{\rm GW}^{\rm max} \lesssim 10^{12}-10^{13}$ Hz for prime sources of cosmological GWBs.

As for the maximal amplitude of the GWB, there is a strong general constraint that applies at all frequencies. It comes from the maximally allowed amount of GW that can be present at the time of Big Bang nucleosynthesis (BBN) and in the cosmic microwave background (CMB) measurements \cite{Kawasaki:1999na,Kawasaki:2000en,Hannestad:2004px,Planck:2018vyg}. It can be written as the bound on the energy-density fraction in GW today $\Omega_{\rm GW}$ spanning over frequency $f_{\rm GW}$ as
\begin{align}
    \int_{f_{\rm GW}^{\rm min}}^{f_{\rm GW}^{\rm max}} \frac{d  f_{\rm GW}}{f_{\rm GW}} \, h^2 \Omega_{\rm GW}(f_{\rm GW}) \lesssim 5.6 \times 10^{-6} \Delta N_{\rm eff}, 
    \label{eq:Neff_bound}
\end{align}
where $h\simeq 0.68$ \cite{Planck:2018jri}, the effective number $\Delta N_{\rm eff} \lesssim 0.2$ of relativistic particle species is bounded by BBN/CMB \cite{Mangano:2011ar, Cyburt:2015mya, Peimbert:2016bdg, Planck:2018jri}, and the $f_{\rm GW}^{\rm min}$ ($f_{\rm GW}^{\rm max}$) is the lower (upper) cutoff frequency of the GWB.
This translates into  $\Omega_{\rm GW} \lesssim 10^{-7}$ for short-lasting sources such as phase transitions, and to $\Omega_{\rm GW} \lesssim 10^{-8}$ for long-lasting sources such as cosmic strings; see Eq.~\eqref{eq:BBN_GW_CS}. 
We compare in Fig.~\ref{fig:money_plot_intro} different types of GW spectra inherited from the early Universe, which would escape detection at present and future interferometers and require UHF-GW experiments. 

The interest in UHF-GW detection has blossomed lately, impulsed by the ``UHF-GW Initiative" \cite{Aggarwal:2020olq}, leading to new ideas for detection techniques,  e.g., \cite{Ballantini:2005am,Arvanitaki:2012cn,Ejlli:2019bqj,Aggarwal:2020umq,LSD:2022mpz,Berlin:2021txa,Berlin:2022hfx,Berlin:2023grv,Goryachev:2021zzn,Goryachev:2014yra,Campbell:2023qbf,Sorge:2023nax,Tobar:2023ksi,Carney:2023nzz,Domcke:2022rgu,Domcke:2023bat,Bringmann:2023gba,Vacalis:2023gdz,Liu:2023mll,Ito:2019wcb,Ito:2020wxi,Ito:2022rxn,Ito:2023bnu}.
Still, this remains extremely challenging experimentally,  and none of the proposals so far reach a sensitivity that enables them to go beyond the BBN bound \eqref{eq:Neff_bound}.
This paper aims to motivate further investigation and provide a concrete science case for UHF-GW detectors: the possibility to probe particle physics at energy scales many orders of magnitude beyond the reach of future particle colliders.

Cosmic strings \cite{Kibble:1976sj, Kibble:1980mv, Hindmarsh:1994re, Vilenkin:2000jqa} are among the most promising sources of GWBs; see   \cite{Auclair:2019wcv, Gouttenoire:2019kij} for recent reviews. Not only do they arise in many well-motivated extensions of the Standard Model of particle physics, but they also scan almost the entire cosmological history. 
A string network evolves into the so-called scaling regime \cite{Kibble:1984hp, Albrecht:1984xv, Bennett:1987vf, Bennett:1989ak, Albrecht:1989mk, Allen:1990tv, Martins:2000cs, Ringeval:2005kr, Vanchurin:2005pa, Martins:2005es, Olum:2006ix, Blanco-Pillado:2011egf, Figueroa:2012kw, Martins:2016ois} where its energy density tracks the total energy density of the Universe, and continuously emits particles and string loops---where the latter subsequently decays into particles or radiates gravitationally \cite{Vachaspati:1984gt,Vilenkin:1986ku,Srednicki:1986xg,Vilenkin:2000jqa,Dabholkar:1989ju,Vilenkin:1982ks}.

With loops being produced throughout cosmological history, the cosmic-string network is a long-lasting source of cosmological GWB, spanning an extremely broad frequency range. It carries information on the cosmic history before BBN, when the age of the Universe is less than $\sim$1 sec, and the energy scale is above $\sim$ MeV and far beyond. This GWB is potentially detectable at planned future GW experiments \cite{Janssen:2014dka, EPTA:2015qep, EPTA:2016ndq, NANOGRAV:2018hou, Weltman:2018zrl, LISA:2017pwj, LISACosmologyWorkingGroup:2022jok, Yagi:2011wg, LIGOScientific:2014qfs, LIGOScientific:2019vic, Hild:2010id, Punturo:2010zz, LIGOScientific:2016wof, EPTA:2023hof, EPTA:2023xxk, NANOGrav:2023hvm, Figueroa:2023zhu, Ellis:2023oxs} and its full spectrum could be reconstructed by space-based and ground-based GW observatories and their synergy \cite{Caprini:2019pxz, Dimitriou:2023knw, Alvey:2023npw}. 
However, we will consider the cosmic-string network decaying well above BBN scale and having the corresponding GWB unobservable at GW detectors below kHz frequencies due to their infrared (IR) cutoff. 
The current bound on the string tension $\mu$ [Eq.~\eqref{eq:string_tension_def}] from pulsar timing arrays ($G\mu \lesssim 10^{-10}$ \cite{NANOGrav:2023hvm, EPTA:2023xxk, Figueroa:2023zhu}),  from LIGO/Virgo/KAGRA ($G\mu \lesssim 10^{-7}$ \cite{LIGOScientific:2021nrg}), and from the CMB temperature anisotropy ($G\mu \lesssim 10^{-7}$ \cite{Planck:2013mgr,Hergt:2016xup}) are evaded in this case. The signals we will consider exceed the largest thermal-plasma GWB \cite{Ringwald:2020ist} assuming the maximal reheating temperature $T_{\rm reh}^{\rm max}$. 

The existence of the IR cutoff is motivated as follows.
A theory leading to cosmic-string formation could be embedded in a theory with a symmetry group larger than $U(1)$ and which undergoes symmetry-breaking multiple times in the early Universe \cite{Lazarides:1981fv,Vilenkin:1982hm,Kibble:1982ae,Vilenkin:1984ib}.
Consider breaking the symmetry groups with the following pattern $G \to H \to K$. When the homotopy group of the final vacuum manifold is trivial, i.e., $\pi_{n}(\mathcal{M}=G/K) = \mathds{1}$, there are no topological defects in the full theory. On the other hand, metastable topological defects can form during each phase transition if the homotopy group of each vacuum manifold ($\mathcal{M}=$ $G/H$ or $H/K$) is non-trivial.

A defect of dimension $p$ is formed when $\pi_{2-p}(\mathcal{M}) \neq \mathds{1}$ where $p=0,1,2$ is for monopole, cosmic string, and domain walls, respectively.
It has been proven that the $p$-dimensional defects must be formed in the phase transition before the formation of those of $(p+1)$-dimension; see e.g., \cite{Vilenkin:2000jqa} and appendix A of \cite{Dunsky:2021tih}. However, the metastable defect of dimension $p$ can be destabilized by the defect of either dimension $p\pm 1$.
The metastable cosmic strings can thus decay via 1) nucleation of monopoles, which are formed in the phase transition before that of the string network, or 2) collapse by attaching to domain walls formed after the string network.

We will discuss, in turn, local and global strings---that result from the breaking of local and global symmetry, respectively---and thus explore the UHF-GW signals from two well-motivated scenarios.
(i) The local cosmic strings decay by monopole-antimonopole pair production \cite{Vilenkin:1982hm, Copeland:1987ht, Preskill:1992ck, Vilenkin:2000jqa, Monin:2008mp, Leblond:2009fq}---which could arise from multiple symmetry breakings in grand unified theories \cite{Lazarides:1981fv, Kibble:1982ae,Vilenkin:1982ks,Vilenkin:1984ib}.
(ii) For global strings, which can be produced in postinflationary axion models, the metastability is automatically built-in via axion domain walls and determined by the axion mass \cite{Vilenkin:1982ks,Hiramatsu:2012gg}.
On the other hand, the high-frequency cutoffs depend on the symmetry-breaking scale when the network is formed, on the small-scale structures (kinks and cusps) of cosmic strings, and on the friction due to string interactions with the thermal plasma.

We briefly recap the cosmic-string GWB from a stable network in section~\ref{sec:recap} and the corresponding ultraviolet (UV) cutoffs in section~\ref{sec:GW_HF_cutoff}. 
Then, section~\ref{sec:local_metastable_network} focuses on the chopped GWB from loops of metastable local strings. Interestingly, we find a large GW signal in the UHF regime, comparable to the BBN bound. In some cases, it exhibits a peak shape as opposed to the usual flat GWB from cosmic strings.
Section~\ref{sec:recontruct} suggests ways to infer information about the underlying microscopic physics of cosmic strings from GW measurements at UHF-GW experiments.
Section~\ref{sec:global_metastable_network} discusses the case of global-string GWB from heavy axions.
Considering the early temporary matter-domination (MD) era induced by decaying axions, we explain why detecting GWB from 
heavy-axion strings would be extremely challenging. We conclude in section~\ref{sec:conclude}.
Appendices contain further details on (i) the effect from the maximal mode of loop oscillation in appendix~\ref{app:kmax_effect}, (ii) the peaked GWB spectrum in appendix~\ref{app:peak_GWB}, (iii) GWB contributions from local-string segments with monopoles on their ends in appendix~\ref{app:segments}, and (iv) the modified causality tail of axion-string GWB by the axion matter domination era in appendix~\ref{app:axion_MD_IR_tail}.

\section{General properties of GWB from cosmic strings}
\label{sec:recap}

We assume the existence of a
complex scalar field $\Phi$, charged under a local or global $U(1)$-symmetry, with the potential 
\begin{align}
V(\Phi) = \lambda(|\Phi|^2 - \eta^2)^2/2.
\label{eq:potential_CS}
\end{align}
The scalar self-coupling $\lambda$ which determines the mass of $\Phi$ can be small in some models where the scalar field is associated with a flat direction in supersymmetric theories \cite{Lazarides:1985ja, Perkins:1998re, Okada:2013vxa, Ellis:2018ojk}. As we shall see below, the scalar self-coupling plays an important role as it determines the temperature when the string network formed and the width of cosmic strings.

Note that this potential represents only one class of the field theories that produce cosmic strings. The potential beyond the quartic type can also lead to cosmic strings as long as the symmetry-breaking pattern allows them (i.e., the first homotopy group of the vacuum manifold is non-trivial). As an example, the nearly-quadratic potential $m^2 |\Phi|^2 \left[2\log\left(|\Phi|/\eta\right)-1\right] + m^2 \eta^2$ which is motivated by theories with gravity-mediated supersymmetry-breaking \cite{Moxhay:1984am} (see a review in appendix D.4 of \cite{Gouttenoire:2021jhk}) can support the cosmic-string formation. With this potential, the formation scale and string's width are controlled directly to the scalar's mass $m$, which can be much smaller than $\eta$.

The $U(1)$-symmetry is preserved at early times and gets spontaneously broken once the temperature drops below $T_{\rm form} \simeq \lambda^{1/2}\eta$ where $\eta$ is the vacuum expectation value of the field. This leads to the formation of the cosmic-string network with the string tension (i.e., energy per unit length) \cite{Vilenkin:2000jqa}
\begin{align}
    \mu \simeq \eta^2 \times \begin{cases}
    1~ ~ &{\rm (local)},\\
    \log(\lambda^{1/2}\eta/H)~ ~ &{\rm (global)},
    \end{cases}
    \label{eq:string_tension_def}
\end{align}
where $H$ is the Universe's expansion rate.
In this work, we remain agnostic on the string formation mechanism (either from thermal effects \cite{Kibble:1976sj, Kibble:1980mv, Hindmarsh:1994re, Vilenkin:2000jqa} or non-perturbative dynamics \cite{Kofman:1995fi, Tkachev:1995md, Kasuya:1997ha, Kasuya:1998td, Tkachev:1998dc}) and scan over the extensive range of string tension $\mu$. 
After the network formation, the string network keeps producing loops. It reaches the scaling regime where its energy density tracks the total energy density of the Universe, $\rho_{\rm net}(t) \simeq \mu/t^2 \simeq G\mu \rho_{\rm tot}(t)$.

The produced loops decay into particles and GW. Local-string loops decay dominantly into GW while global-string loops decay dominantly into Goldstone radiation.
The energy-density spectrum of GWB can be written as a superposition of many loop populations producing GW at time $\tilde{t}$ and of many oscillation $k^{\rm th}$-modes,
\begin{align}
    \Omega_{\rm GW}(f_{\rm GW}) = &\frac{1}{\rho_{c,0}}\sum_{k = 1}^{k_{\rm max}}\frac{2k}{f_{\rm GW}} \cdot \Gamma^{(k)}G\mu^2 \nonumber\\
    &\times \int_{t_{\rm form}}^{t_0} n_{\rm loop}(\tilde{t})\left[\frac{a(\tilde{t})}{a(t_0)}\right]^5 d \tilde{t},
    \label{eq:GWB_spectrum_master}
\end{align}
where $\rho_{c,0}$ is the Universe's energy density today, the GW emission efficiency per mode is $\Gamma^{(k)} \simeq \Gamma k^{-4/3}/\zeta(4/3)$ with $\zeta$ being the Riemann zeta function and $\Gamma \simeq 50$ \cite{Blanco-Pillado:2017oxo} being the total efficiency (the number 4/3 is used for loops with cusps), $n_{\rm loop}$ is the number density of loops, $\tilde{t}$ is the GW emission time, and $t_0$ is the time today. Eq.~\eqref{eq:GWB_spectrum_master} sums the number of modes up to $k_{\rm max}$, which can be treated as infinity; the effect of finite $k_{\rm max}$ is present at the high-frequency end of the spectrum, as discussed in appendix~\ref{app:kmax_effect}.

All GWB spectra in this work are calculated numerically by following \cite{Gouttenoire:2019kij} (see also \cite{Cui:2017ufi,Cui:2018rwi,Chang:2019mza,Chang:2021afa}) where $n_{\rm loop}$ relies on the velocity-dependent one-scale (VOS) model \cite{Martins:1995tg,Martins:1996jp, Martins:2000cs, Sousa:2013aaa, Sousa:2014gka, Correia:2019bdl}.
That is,
\begin{align}
    n_{\rm loop}(t) = (0.1) \frac{C_{\rm eff}(t_i)}{\alpha (\alpha + \Gamma G\mu + \Gamma_{\rm gold}) t_i^4} \left[\frac{a(t_i)}{a(\tilde{t})}\right]^3,
    \label{eq:loop_numberdensity_stable}
\end{align}
where $\alpha \sim \mathcal{O}(0.1)$ \cite{Blanco-Pillado:2013qja} is the initial loop size as a fraction of Hubble horizon $H^{-1}$, the prefactor (0.1) means only 10\% of loops contributed to GWB \cite{Blanco-Pillado:2013qja}, $\Gamma_{\rm gold}\simeq 0$ for local strings and $\simeq 65/[2\pi \log(\eta t)]$ for global strings \cite{Vilenkin:1986ku} is the loop-length shrinking rate by emitting Goldstone bosons, $t_i$ is the loop formation time [which can be written in terms of $\tilde{t}$ and $f_{\rm GW}$ using Eq.~\eqref{eq:loop_length_evo}], and $C_{\rm eff}$ is the loop production coefficient which is solved from the VOS equations; see e.g., in \cite{Gouttenoire:2019kij}, section~4 for local strings and appendix~F for global strings. 
This work uses the input for the VOS equations from Nambu-Goto simulation \cite{Martins:2000cs}, although the small $\lambda$ might change the evolution of the string network (e.g., loop formation and particle production).

 In the high-frequency regime---corresponding to loop produced and emitting GW deep inside the radiation-dominated Universe, the amplitude \eqref{eq:GWB_spectrum_master} reads (see derivation in  \cite{Gouttenoire:2019kij,Gouttenoire:2021jhk})
\begin{align}
    h^2 \Omega_{\rm GW} \simeq \begin{cases}
    {\rm (local)} ~ 1.5 \cdot 10^{-10} \mathcal{G}[T(f_{\rm GW})] \\
    ~ ~ \times \left[\frac{G \mu}{10^{-11}}\right]^{\frac{1}{2}} \left[\frac{\alpha}{0.1}\right]^{\frac{1}{2}} \left[\frac{50}{ \Gamma}\right]^{\frac{1}{2}},
    \\[0.5em]
    {\rm (global)} ~ 1.6 \cdot 10^{-11} \left(\frac{\eta}{10^{15} \, \rm GeV}\right)^4 \\
~ ~ \times \mathcal{G}[T(f_{\rm GW})]\left[\frac{\mathcal{D}(\eta, f_{\rm GW})}{94.9}\right]^3 \left[\frac{C_{\rm eff}(f_{\rm GW})}{2.24}\right],
    \end{cases}
    \label{eq:GWB_stable}
\end{align}
where $\mathcal{G}(T) \equiv [g_*(T)/g_*(T_0)][g_{*s}(T_0)/g_{*s}(T)]^{4/3}$ with $g_*$ ($g_{*s}$) the relativistic degrees of freedom in energy (entropy) density (taken from \cite{Saikawa:2018rcs}), and $T_0$ the photon temperature today. 
The log-correction is 
\begin{align}
    \mathcal{D}(\eta, f_{\rm GW})=\log\left[5.7 \cdot 10^{18} \left(\frac{\eta}{10^{15} \, \rm GeV}\right) \left(\frac{1 \, \rm kHz}{f_{\rm GW}}\right)^2\right].
    \label{eq:log_dependence_D_factor}
\end{align}
This work uses the exponent `3' of the log-dependent term $\mathcal{D}$ (similar to \cite{Hindmarsh:2021vih}), although this is still under debate as the exponent `4' is found in some simulation results \cite{Gorghetto:2018myk,Gorghetto:2020qws,Gorghetto:2021fsn}.
The final $\Omega_{\rm GW}$ in the latter case could be enhanced from our result due to the log factor by $\mathcal{O}(100)$.
Note also that a potentially less efficient GW emission from a single loop was found in \cite{Baeza-Ballesteros:2023say}, which suggests that $\Omega_{\rm GW}$ would be weaker by  $\mathcal{O}(10^{4})$ compared to our result.

\begin{figure}[t!]
\includegraphics[width=\linewidth]{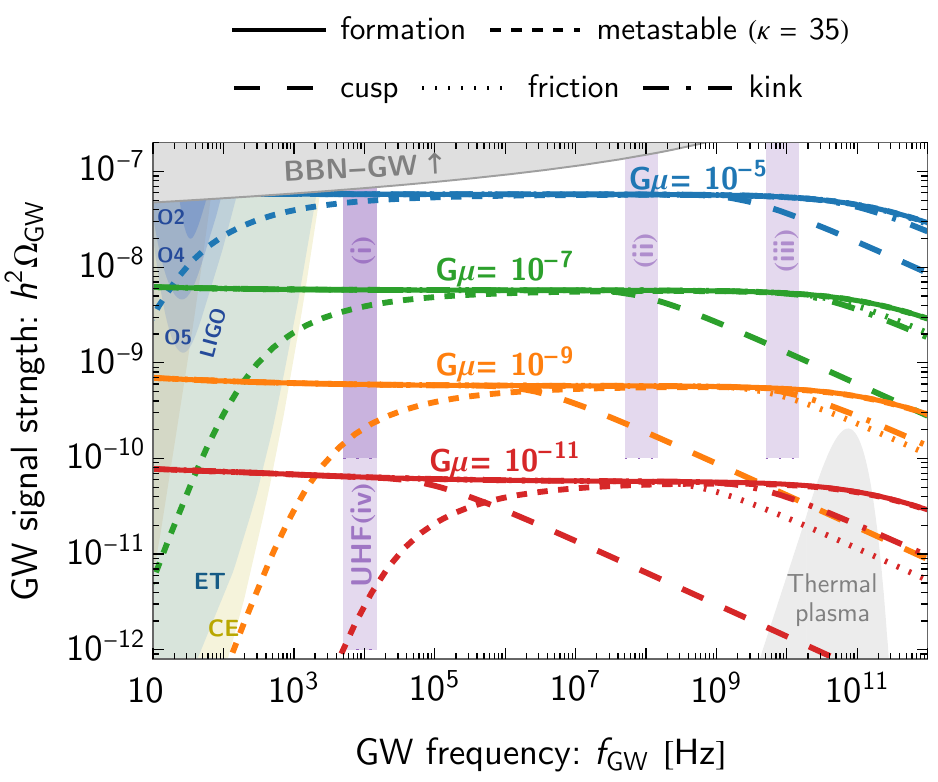}\\[-1em]
\caption{Local-string GWBs in the UHF range featuring different cutoffs. The solid lines show the stable-string GWB with the formation cutoff \eqref{eq:formation_cut_off}. The dashed,  dotted, and dot-dashed lines correspond to the cutoffs from cusp \eqref{eq:freq_cusp}, kink \eqref{eq:freq_kink}, and friction \eqref{eq:freq_friction}, respectively, assuming the stable-string network. The four rectangular purple regions denote hypothetical sensitivities of four fictional UHF-GW experiments; see Eq.~\eqref{eq:UHF_experiment}. The ``thermal plasma" gray region is the GWB predicted in \cite{Ghiglieri:2015nfa, Ghiglieri:2020mhm, Ringwald:2020ist}, assuming the maximal reheating temperature $T_{\rm reh}^{\rm max} \simeq 6 \times 10^{15} ~ {\rm GeV}$ \cite{Ghiglieri:2015nfa, Ghiglieri:2020mhm, Ringwald:2020ist}, while the upper ``\textsc{bbn-gw}" gray region is excluded by the BBN bound \eqref{eq:Neff_bound}.}
\label{fig:local_GWB_spectrum}
\end{figure}

The GWB from local strings is $f_{\rm GW}$-independent, while the global-string GWB is log-suppressed at high frequencies.
We show the GWB spectra from local strings in  Fig.~\ref{fig:local_GWB_spectrum}, where the  IR and UV slopes are explained in the next sections. It is clear from this figure that large signals touching the BBN bound can arise, associated with $G\mu$ approaching $10^{-5}$ and thus a scale of  $U(1)$ symmetry breaking close to $10^{16}$ GeV.
We do not show the GWB spectra from global strings. As we shall see below, the metastability of heavy-axion strings comes with an early axion-matter-dominated era that dilutes and heavily suppresses the GWB.

The broadband GWB spectrum is the result of the superposition of  GW generated by many populations of loops produced at different temperatures. 
Each emits GW at frequency $f_{\rm GW}^{\rm emit} \simeq 2k/l$ \cite{Kibble:1982cb, Hindmarsh:1994re} where $k$ is the mode number of loop oscillation and the loop's size is
\begin{align}
l(t,t_i) = \alpha t_i - (\Gamma G\mu + \Gamma_{\rm gold}) (t-t_i).
\label{eq:loop_length_evo}
\end{align}
For the loop population created at temperature $T$, the GWB is sourced maximally at frequency today, $f_{\rm GW} = f_{\rm GW}^{\rm emit}(t_{\rm emit},t_i(T)) [a(t_{\rm emit})/a_0]$. As shown in \cite{Gouttenoire:2019kij,Gouttenoire:2021jhk} (see also \cite{Cui:2017ufi,Cui:2018rwi,Chang:2019mza,Chang:2021afa}), the GWB today's frequency and Universe's temperature relation can be written respectively for local and global strings as
\begin{align}
    f_{\rm GW}(T) \simeq \begin{cases} {\rm(local)} ~ 2 ~ {\rm kHz} \left[\frac{T}{10^6 ~ \rm GeV}\right] \left[\frac{10^{-11}}{G\mu}\right]^{\frac{1}{2}} \left[\frac{g_*(T)}{g_*(T_0)}\right]^{\frac{1}{4}},\\[0.5em]
    {\rm(global)} ~ 1.1 ~ {\rm kHz}  \left[\frac{T}{10^{8} ~ \rm GeV}\right] \left[\frac{g_*(T)}{106.75}\right]^{\frac{1}{4}}.
    \end{cases}
    \label{eq:freq_temp_relation}
\end{align}
Note that the above frequency depends on $G\mu$ for local strings because the GW is sourced maximally around the time \cite{Cui:2018rwi,Gouttenoire:2019kij}
\begin{align}
\label{eq:ttilde}
\tilde{t} \simeq \alpha/(2 \Gamma G\mu) t_i.
\end{align}
In contrast, global strings quickly emit GW after loop production $\tilde{t} \sim t_i$.

Eq.~\eqref{eq:freq_temp_relation} indicates that the GW contribution at higher frequencies comes from smaller loops produced at higher energy scales where microscopic properties play a more prominent role. 
Ultimately, the GW emission, which relies on the collective motion of the smaller loops, is more suppressed \cite{Brandenberger:1986vj, Blanco-Pillado:1998tyu}.
We now recap the different types of UV cutoffs in the next section; see also \cite{Auclair:2019jip} for a review. 

\section{High-frequency cutoffs of cosmic-string GWB }
\label{sec:GW_HF_cutoff}

A string-loop population contributes maximally to the GWB at the frequency in Eq.~\eqref{eq:freq_temp_relation}. It generates the UV tail with a slope of $\Omega_{\rm GW} \propto f_{\rm GW}^{-1}$ for a single proper loop-oscillation mode, both local and global strings; see \cite{Cui:2018rwi, Gouttenoire:2019kij, Simakachorn:2022yjy, Ghoshal:2023sfa}. 
By summing over large harmonics, the UV slope changes from $-1$ to $-1/3$ in the case of loops with cusps; see derivation, e.g., in \cite{Blasi:2020wpy, Gouttenoire:2019kij}.
We show in appendix~\ref{app:kmax_effect} and Fig.~\ref{fig:app_kmax} that the precise calculation involving the $k$-dependent cutoff (due to the string's width) leads to a slight modification in the slope above the UV cutoff and does not change the position of the cutoff (for $G\mu \lesssim 10^{-4}$).
However, this is computationally expensive; therefore, in all figures, we apply that the spectra fall as $f_{\rm GW}^{-1/3}$ beyond the UV cutoffs, which we will now discuss.

\subsection{Formation cut-off} 
The most conservative UV cutoff comes from the energy scale of the string-network formation.
Using Eq.~\eqref{eq:freq_temp_relation} with $T_{\rm form} \simeq \lambda^{1/2}$ $ \eta \simeq \lambda^{1/2}$ $m_{\rm Pl}\sqrt{G\mu}$ where $m_{\rm Pl}=\sqrt{1/G}$ is the Planck mass, we obtain the formation cutoff for local strings,
\begin{align}
    \boxed{f_{\rm GW}^{\rm form} \simeq \begin{cases} 
    {\rm \small (local)} ~ 182 ~ {\rm GHz}  \, \sqrt{\lambda} \left[\frac{g_*(T_{\rm form})}{106.75}\right]^{\frac{1}{4}},\\[0.5em]
    {\rm \small (global)} ~ 11 ~ {\rm GHz} \, \sqrt{\lambda} \left[\frac{\eta}{10^{15} ~ \rm GeV}\right] \left[\frac{g_*(T_{\rm form})}{106.75}\right]^{\frac{1}{4}}.
    \end{cases}} 
    \label{eq:formation_cut_off}
\end{align}
We set $\lambda = 1$, except for section ~\ref{sec:recontruct}.
For local strings, the formation cutoff does not depend on the string tension, as a higher GW frequency at production from an earlier network formation is compensated by the longer loop lifetime $\tilde{t}$. For global strings, the formation cutoff is proportional to the symmetry-breaking scale $\eta$, reflecting the fast decay of string loops.

We will treat $\eta$ as a free parameter in this work. Its precise range depends on the cosmic-string formation mechanisms. For example, the $U(1)$ symmetry can be restored by quantum fluctuations during inflation, leading to cosmic-string formation at the end of inflation \cite{Bunch:1978yq, Linde:1983mro, Starobinsky:1994bd}. This can happen for $\eta < H_{\rm inf}/(2\pi) \lesssim 9.7 \times 10^{12} ~ {\rm GeV}$ where the last inequality uses the current CMB bound \cite{Planck:2018jri}.
There are other well-motivated mechanisms which generate postinflationary cosmic strings with $\eta \gg H_{\rm inf}$ (cf. also appendix H of \cite{Gorghetto:2021fsn}): (i) non-perturbative effects during preheating  \cite{Kofman:1995fi, Tkachev:1995md, Kasuya:1997ha, Kasuya:1998td, Tkachev:1998dc}, (ii) thermal effects due to the interaction of the complex scalar field $\Phi$ with the thermal plasma which allows for $T_{\rm form} \simeq \lambda^{1/2}\eta < T_{\rm reh}$ where the reheating temperature can be as large as $T_{\rm reh} \lesssim 6 \times 10^{15} ~ {\rm GeV}$ \cite{Planck:2018jri}, and (iii) the large and positive Hubble-dependent mass induced by the coupling with the inflaton field\footnote{The inflaton $\chi$ could trap $|\Phi| \to 0$ during inflation via the effective mass term arising from the coupling $\mathcal{L} \supset |\Phi|^2 \chi^2$.}.

\subsection{Small-scale structures: kinks and cusps}
The loops produced early are smaller and poorly approximated by the linelike description.
Especially near small-scale structures such as cusps and kinks, they can dominantly decay via radiating non-perturbatively massive particles instead of emitting GW \cite{Brandenberger:1986vj,Blanco-Pillado:1998tyu, Olum:1998ag,Matsunami:2019fss, Blanco-Pillado:2023sap}. Global string loops decay shortly after their formation such that a kink-collision or a cusp does not have time to develop.
We, therefore, would not expect a cutoff on the global-string GWB spectrum from the cusp or kink, although this needs to be confirmed with numerical field-theory simulations. In this subsection,  the formulae for cusp/kink cutoffs apply to  local strings only.

The massive radiation is more efficient than the GW emission ($P_{\rm GW} = \Gamma G\mu$) when the loop length is smaller than $l_m = \beta_m {\mu^{-1/2}}/{\left(\Gamma G \mu\right)^m}$
where $m=1,~2$ for kink- and cusp-dominated loops, respectively \cite{Auclair:2019jip}. The factor $\beta_m$ depends on the detailed small-scale structure of the string loop and typically grows with the number of kinks and cusps
or the self-coupling $\lambda$.
The overlapping cusp segment of length $\sqrt{wl}$ with the string width $w \simeq m_\Phi^{-1} \simeq (\lambda^{1/2} \eta)^{-1}$ annihilates into particles \cite{Brandenberger:1986vj,Blanco-Pillado:1998tyu, Olum:1998ag}. The emission power per oscillation period $l/2$ is $P_{c} \sim N_c \mu \sqrt{wl}/l \simeq N_c \lambda^{-1/4}\mu^{3/4}/l^{1/2}$ with  $N_c$ is the number of cusps which is typically two in each oscillation \cite{Blanco-Pillado:2015ana}.
The energy emission from kinks is $P_{k} \simeq N_{kk} \mu w/l \simeq N_{kk} \lambda^{-1/2} \mu^{1/2}/l$ with $N_{kk}$ is the number of kink-kink oscillation, which in some models can be as large as $N_{kk} \sim \mathcal{O}(10^3-10^{6})$ 
\cite{Ringeval:2017eww, Binetruy:2010bq, Binetruy:2010cc}.
Thus, we have
\begin{align}
    \beta_c \simeq N_c \lambda^{-1/4}, ~ {\rm and} ~ \beta_k \simeq N_{kk} \lambda^{-1/2}.
    \label{eq:beta_cusp_kink}
\end{align}

The loops formed with a size $\alpha t_i$ smaller than $l_m$, or equivalently, formed above temperature
\begin{align}
    T_{\rm cusp} &\simeq 0.59 ~  {\rm PeV} \sqrt{\frac{1}{\beta_c}}\left(\frac{G\mu}{10^{-11}}\right)^{\frac{5}{4}} \left[\frac{106.75}{g_*(T)}\right]^{\frac{1}{4}},
    \label{eq:temperature_cusp}\\
    T_{\rm kink} &\simeq 26.2 ~ {\rm EeV} \sqrt{\frac{1}{\beta_k}}\left(\frac{G\mu}{10^{-11}}\right)^{\frac{3}{4}} \left[\frac{106.75}{g_*(T)}\right]^{\frac{1}{4}},\label{eq:temperature_kink}
\end{align}
decay into massive particles with mass of order $\eta$ and should not contribute to the GWB \cite{Auclair:2019jip}. 
We assume these subsequently decay into  Standard Model particles and do not lead to additional constraints.
Using Eq.~\eqref{eq:freq_temp_relation} (where factor 2 is replaced by 45 not to include the later-time network evolution \cite{Gouttenoire:2019kij}), the GWB spectrum has a high-frequency cutoff due to either kinks or cusps at \cite{Gouttenoire:2019kij}

\begin{align}
    \boxed{f_{\rm GW}^{\rm cusp} \simeq 62.3 ~ {\rm kHz} \sqrt{\frac{1}{\beta_c}}\left(\frac{G\mu}{10^{-11}}\right)^{3/4},}
    \label{eq:freq_cusp}
\end{align}
\begin{align}
    \boxed{f_{\rm GW}^{\rm kink} \simeq 2.79 ~ {\rm GHz} \sqrt{\frac{1}{\beta_k}}\left(\frac{G\mu}{10^{-11}}\right)^{1/4}.}
    \label{eq:freq_kink}
\end{align}
Using Eqs.~\eqref{eq:GWB_stable}, \eqref{eq:freq_cusp} and \eqref{eq:freq_kink} and varying $G\mu$, the  GW amplitude at the cutoffs are
\begin{align}
    h^2 \Omega_{\rm GW}^{\rm kink} &\simeq 7.35 \times 10^{-12} ~\beta_k \left(\frac{f_{\rm GW}^{\rm kink}}{1~{\rm GHz}}\right)^{2},\label{eq:kink_cutoff_position}\\
    h^2 \Omega_{\rm GW}^{\rm cusp} &\simeq 3.69 \times 10^{-10} ~\beta_c^{\frac{1}{3}}\left(\frac{f_{\rm GW}^{\rm cusp}}{1~{\rm MHz}}\right)^{\frac{2}{3}}.\label{eq:cusp_cutoff_position}
\end{align}

\subsection{Thermal friction}
Cosmic strings can experience a frictional force on top of the Hubble expansion if they interact with other particles of the thermal plasma.
This friction affects the long-string evolution and also the oscillation of loops.
The length scale for the efficient thermal friction is given by $l_{\rm fric} \sim \mu/(\sigma \rho_{\rm th}) \equiv \mu/(\beta_{\rm fric} T^3)$, where the scattering cross-section per unit length $\sigma \equiv \beta_{\rm fric}/T$ with $\beta_{\rm fric}$ depending on the nature of interactions \cite{Vilenkin:1991zk, Vilenkin:2000jqa}, and the energy density of thermal plasma $\rho_{\rm th} \sim T^4$.  
For example, the Aharonov-Bohm interaction induces friction with $\beta_{\rm fric} \sim 1$ \cite{Aharonov:1959fk, Vilenkin:1991zk, Vilenkin:2000jqa}.
Thermal friction is efficient at high $T$ until $2H \gtrsim l_{\rm fric}^{-1}$ when the temperature drops below
\begin{align}
    T_{\rm fric} \simeq 4.19 ~ {\rm EeV} ~ \beta_{\rm fric}^{-1} \left(\frac{G\mu}{10^{-11}}\right) \left[\frac{g_*(T)}{106.75}\right]^{\frac{1}{2}}.
    \label{eq:temperature_friction}
\end{align}
For local strings, this is associated with the frequency [using Eq.~\eqref{eq:freq_temp_relation}]
\begin{align}
    \boxed{f_{\rm GW}^{\rm fric} \simeq 0.45 ~ {\rm GHz} ~ \beta_{\rm fric}^{-1} \left(\frac{G\mu}{10^{-11}}\right)^{\frac{1}{2}} \left[\frac{g_*(T)}{106.75}\right]^{\frac{3}{4}},}
    \label{eq:freq_friction}
\end{align}
and the GW amplitude \eqref{eq:GWB_stable}, when varying $G\mu$,
\begin{align}
    h^2 \Omega_{\rm GW}^{\rm fric} &\simeq 1.26 \times 10^{-10} ~\beta_{\rm fric} \left(\frac{f_{\rm GW}^{\rm fric}}{1~{\rm GHz}}\right).\label{eq:friction_cutoff_position}
\end{align}
The friction cutoff would carry information about the scalar field couplings to particles in the plasma. 
We will not discuss it for global strings since, as we will see, the UHF GWB from global axionic strings is not observable. In general, the GWB from cosmic strings is observable only at large string scales $\eta$ (corresponding to the axion decay constant $f_a$), which is severely constrained for the light axion \cite{Chang:2021afa,Gorghetto:2021fsn,Gorghetto:2023vqu, Servant:2023mwt}. For very heavy axions, we will show that a matter-domination era is induced, which suppresses the GWB further; see section~\ref{sec:global_metastable_network}.

UHF-GW experiments with sensitivity a few orders of magnitude below the BBN-GW bound would probe the nature of field theory at high energy scales. For example, as illustrated in Fig.~\ref{fig:local_GWB_spectrum}, an experiment operating at $0.1$ GHz with $\Omega_{\rm sens}h^2 \sim 10^{-8}$ (or characteristic strain $h_c \sim 10^{-30}$) would be able to probe the cusp cutoff of GWB from grand-unified-theory strings with $G\mu \sim [10^{-6},10^{-7}]$ or $\eta \sim [4 \cdot 10^{15}, 1.2 \cdot 10^{16}]$ GeV. 
Cosmic strings associated with such high values of $G\mu$ can be compatible with constraints from sub-kHz GW experiments if they are metastable \cite{Vilenkin:1982hm, Preskill:1992ck, Vilenkin:2000jqa, Monin:2008mp, Leblond:2009fq, Dror:2019syi, Buchmuller:2019gfy, Buchmuller:2021mbb,Buchmuller:2023aus}. In the next section, we discuss metastable local strings, and in Section \ref{sec:global_metastable_network}, we discuss metastable global strings.

\section{Metastable local-string GWB: IR cutoff from monopoles}
\label{sec:local_metastable_network}

We consider the local symmetry-breaking pattern giving metastable cosmic strings that are destabilized by monopoles. We will remain agnostic about the UV completions; for examples of such theories and their GWB production, see \cite{Buchmuller:2021mbb,Buchmuller:2023aus}. 
As discussed in the introduction, the formation of monopoles at energy scale $m_M$ has to happen at the phase transition before the string formation at scale $T_{\rm form}$.
A period of inflation at scale $E_{\rm inf}$ between monopole and string formations is required to dilute the monopoles away\footnote{Without inflation, the formation of cosmic strings can solve the monopole problem, but the cosmic strings will be short-lived and do not strongly source the GWB \cite{Dunsky:2021tih}.}. So, the cosmic strings can produce the GWB long before the nucleation of monopoles induces the breaking of the string network and its decay at temperature $T_{\rm brk}$. Therefore, the sequence of events follows the order of energy scales
\begin{align}
    m_M > E_{\rm inf} > T_{\rm form} > T_{\rm brk}.
\end{align}

\subsection{Metastability cutoff}

The metastability of the string network is induced by the nucleation of a monopole-antimonopole pair on strings via quantum tunneling with the rate per unit length \cite{Preskill:1992ck,Monin:2008mp,Leblond:2009fq}
\begin{align}
    \Gamma_d = \frac{\mu}{2\pi}e^{-\pi \kappa} ~ ~ {\rm with} ~ ~ \sqrt{\kappa} \equiv \frac{m_M}{\eta}.
\end{align}
The ratio $\kappa$ is model-dependent; see for example \cite{Masoud:2021prr, Buchmuller:2023aus}.
The expression for $\Gamma_d$ that will be used in this work assumes infinitesimally small monopoles and strings. It was recently found that the finite size of the defects leads to a $\mathcal{O}(1)$-correction of $\kappa$ which highly depends on the UV completion of the theory \cite{Chitose:2023dam} (based on the unwinding string ansatz \cite{Shifman:2002yi}). Besides, more corrections could arise for  $\kappa \sim \mathcal{O}(1)$ when the strings and monopoles have comparable sizes. Their dynamics need to be studied via numerical simulation and are beyond the scope of this work.

In this paper, we treat $\kappa$ as a free parameter.
The breakage of strings happens when $\Gamma_d \ell(t_{\rm brk}) = H(t_{\rm brk})$. Using $\ell(t_{\rm brk}) \simeq  t_{\rm brk}$ and
$H\sim 1/t$, we have 
\begin{align}
    t_{\rm brk} \simeq   \Gamma_d^{-\frac{1}{2}},
    \label{eq:t_brk}
\end{align}
or equivalently in the radiation era ($H= \frac{T^2}{M_{\rm Pl}}\sqrt{\frac{\pi^2}{90} g_*}$ where $M_{\rm Pl} = (8\pi G)^{-1/2}$ is the reduced Planck mass),
\begin{align}
    T_{\rm brk} \simeq 1.74 ~ {\rm TeV} \left[\frac{G\mu}{10^{-11}}\right]^{\frac{1}{4}} \left[\frac{106.75}{g_*(T)}\right]^{\frac{1}{4}} e^{-\frac{\pi}{4}(\kappa-36)},
    \label{eq:metastable_breaking_temperature}
\end{align}

\begin{figure*}[t!]
{\bf Metastable local-string GWB}\\[-0.5em]
\includegraphics[width=0.495\linewidth]{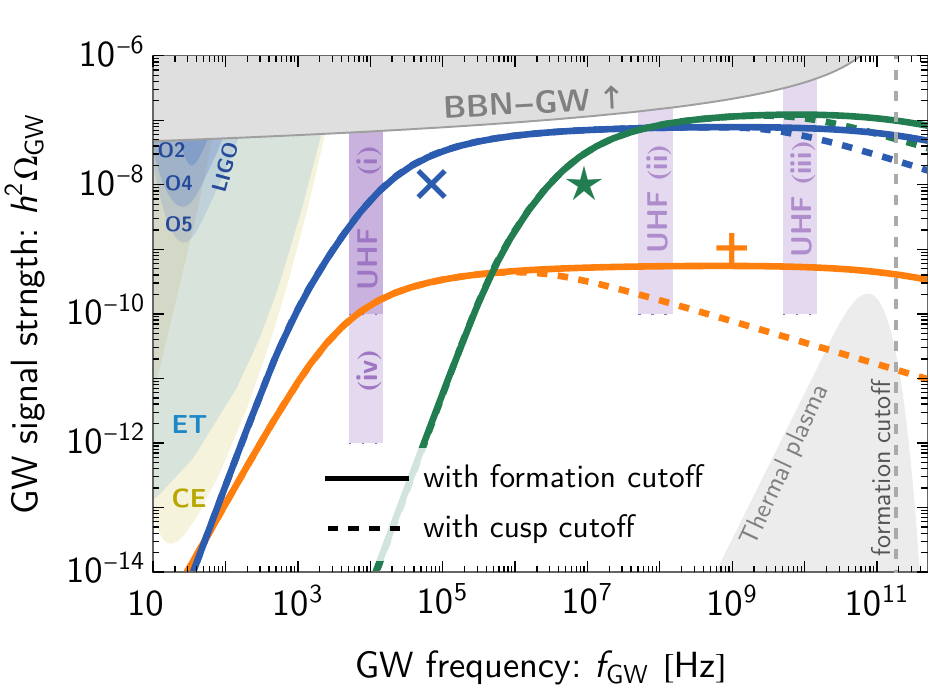}\hfill
\includegraphics[width=0.495\linewidth]{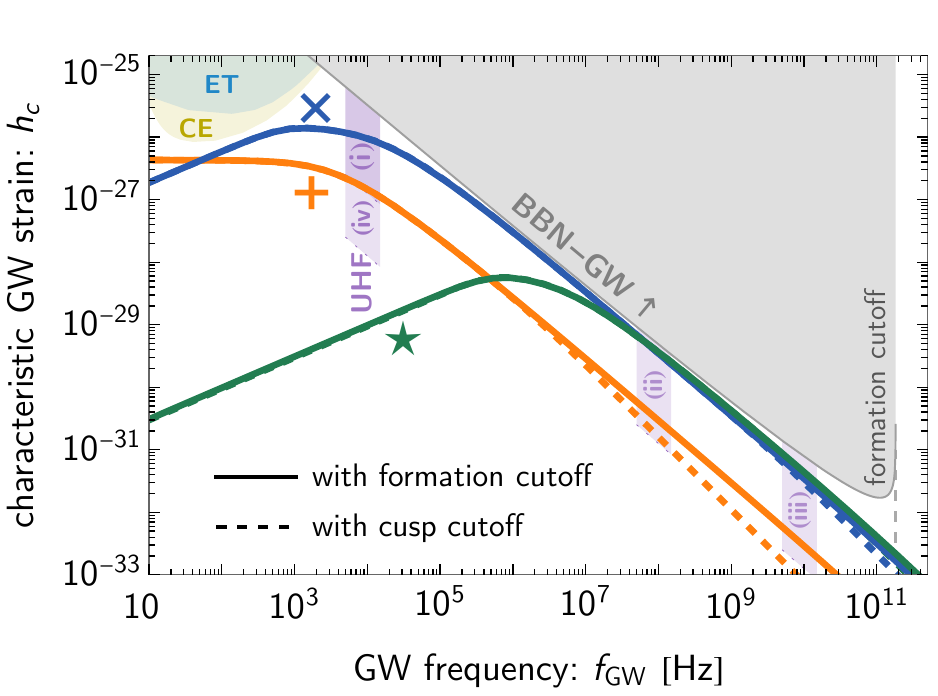}\\[-1em]
\caption{GWB spectra from metastable cosmic strings with the tension $G\mu$ and the metastability parameter $\kappa$ given by the three benchmark values in Fig.~\ref{fig:local_GW_GWB_contour}.
The solid lines assume the GWB with the formation cutoff indicated by the vertical dashed gray line, while the dashed lines assume the GWB with cusp cutoffs.
The IR tails of the spectra exhibit the transition from $f_{\rm GW}^2$ to $f_{\rm GW}^3$ in the low-frequency direction; see Eq.~\eqref{eq:causality_turning_point}.
The top gray region denotes the BBN bound in Eq.~\eqref{eq:BBN_GW_CS}.
The bottom gray region is the thermal plasma GWB. The purple rectangles denote hypothetical GW experiments operating at the UHF range [defined by  Eq.~\eqref{eq:UHF_experiment}], while other colored regions are the power-law integrated sensitivity curves of current and future GW observatories, taken from \cite{Gouttenoire:2019kij}.}
\label{fig:local_GW_GWB}
\end{figure*}

\begin{figure*}[t!]
\includegraphics[width=0.495\linewidth]{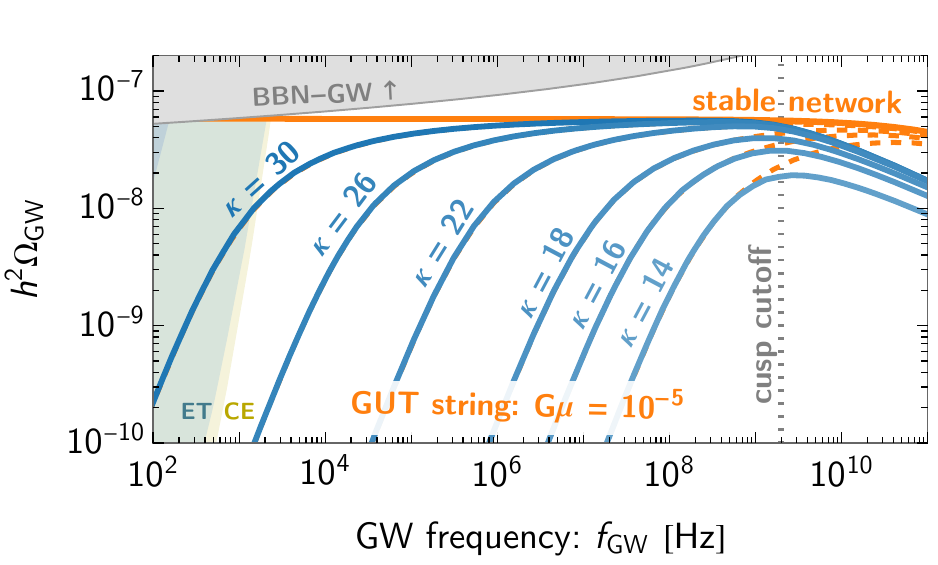}
\hfill
\includegraphics[width=0.495\linewidth]{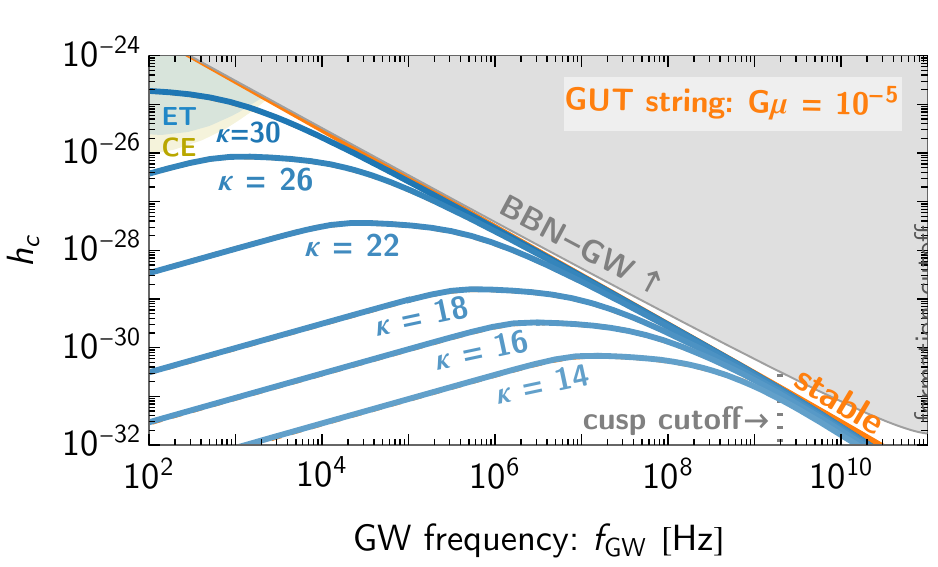}\\[-1em]
\caption{Local-string GWBs in blue with the metastability and cusp cutoffs, at fixed $G\mu=10^{-5}$, for different $\kappa$ where the dashed orange lines assume the cutoff from formation instead of the cusp.
The spectrum exhibits a flat plateau shape when $f_{\rm GW}^{\rm meta} < f_{\rm GW} < f_{\rm GW}^{\rm cusp}$.
For $f_{\rm GW}^{\rm meta} > f_{\rm GW}^{\rm cusp}$, the spectrum has a peak shape with the suppressed amplitude depending on the ratio $f_{\rm GW}^{\rm meta}/f_{\rm GW}^{\rm cusp}$ [see Eq.~\eqref{eq:GW_suppressed_two_cutoffs}].}
\label{fig:local_GW_GWB_combined_cutoff}
\end{figure*}

Monopoles have two effects: (i) terminating loop production after $t_{\rm brk}$ and (ii) suppressing the number density of the existing loops. 
The modified number density of loops is \cite{Buchmuller:2021mbb,Buchmuller:2023aus}
\begin{align}
    n_{\rm loop}^{\rm brk} = n_{\rm loop}^{\rm stable} \Theta(t_{\rm brk} - t_i) \mathcal{E}(l,t),
    \label{eq:local_loop_density}
\end{align}
where $n_{\rm loop}^{\rm stable}$ is the loop number density of the stable network in Eq.~\eqref{eq:loop_numberdensity_stable}, $t_i$ is the time of loop production, and the exponential suppression of the loop number density---as they break into monopoles over time---is \cite{Leblond:2009fq, Buchmuller:2021mbb,Buchmuller:2023aus}
\begin{align}
    \mathcal{E}(l,t) = e^{-\Gamma_d\left[l(t)(t-t_{\rm brk}) + \frac{1}{2}\Gamma G\mu\left(t-t_{\rm brk}\right)^2\right]}.
    \label{eq:suppression_factor}
\end{align}
The GWB from these decaying loops is calculated from Eq.~\eqref{eq:GWB_spectrum_master} using the loop number density $n_{\rm loop}^{\rm brk}$ and is shown in Figs.~\ref{fig:money_plot_intro}, \ref{fig:local_GWB_spectrum}, \ref{fig:local_GW_GWB}, \ref{fig:local_GW_GWB_combined_cutoff}, and \ref{fig:money_plots} (including Fig.~\ref{fig:local_GW_GWB_combined_cutoff_app}).

From Eq.~\eqref{eq:suppression_factor}, one derives that the suppression of loop number density takes effect at the time 
\begin{align}
    t_{\rm sup} \simeq \sqrt{2/ (\Gamma_d \Gamma G\mu)}.
    \label{eq:t_sup}
\end{align}
One might mistakenly associate the IR cutoff to the time $t_{\rm brk}$.
In fact, the first population of loops that experiences number-density suppression and hence the suppressed GWB are loops contributing to GWB maximally at time $t_{\rm sup}$.
These loops are produced at the time $t_i = 2 \Gamma G \mu t_{\rm sup}/\alpha$, using $\tilde{t} = t_{\rm sup}$ and Eq.~\eqref{eq:ttilde}. We arrive at the temperature,
\begin{align}
    T_{i, \rm sup} \simeq 4.9 \times 10^{4} ~ {\rm GeV} \left[\frac{106.75}{g_*(T_{\rm sup})}\right]^{\frac{1}{4}} e^{-\frac{\pi}{4}(\kappa - 36)},
    \label{eq:metastability_GW_suppression_temperature}
\end{align}
Using Eq.~\eqref{eq:freq_temp_relation} with $T_{i,\rm sup}$, the IR cutoff on the metastable-string GWB is at the frequency
\begin{align}
    \boxed{f_{\rm GW}^{\rm meta} \simeq 0.23 ~ {\rm MHz} \left(\frac{10^{-11}}{G\mu}\right)^{\frac{1}{2}} e^{-\frac{\pi}{4}(\kappa - 36)},}
    \label{eq:metastable_cutoff_freq}
\end{align}
where we multiply by the factor of $10^3$ to match the spectra obtained numerically. 
This accounts for non-linear $t$-dependence of $\mathcal{E}$ and the higher-mode summation such that the cutoff is defined where the amplitude drops by $\sim$50\% from the stable-string prediction. For this reason, our Eq.~\eqref{eq:metastable_cutoff_freq} is different from Eq.~(33) of \cite{Buchmuller:2021mbb} by $\mathcal{O}(10^2)$.
The GWB spectrum changes from a flat shape for $f_{\rm GW} \gtrsim f_{\rm GW}^{\rm meta}$ to $\Omega_{\rm GW} \propto f_{\rm GW}^2$ for $f_{\rm GW} \lesssim f_{\rm GW}^{\rm meta}$ \cite{Buchmuller:2021mbb}.
When varying $G\mu$, the amplitude at the metastability cutoff follows
\begin{align}
    h^2 \Omega_{\rm GW}^{\rm meta} &\simeq 1.33 \times 10^{-11} \left(\frac{1~{\rm MHz}}{f_{\rm GW}^{\rm meta}}\right) e^{-\frac{\pi}{4}(\kappa - 36)}.\label{eq:local_metastability_cutoff_position}
\end{align}

At frequency lower than the IR cutoff $f_{\rm GW}^{\rm meta}$, there is another characteristic scale corresponding to the GWB contribution from the last loop population formed at $T_{\rm brk}$ in Eq.~\eqref{eq:metastable_breaking_temperature}; this is at frequency \begin{align}
    \boxed{f_{\rm GW}^{\rm brk} \simeq 25 ~ {\rm mHz} \left(\frac{G\mu}{10^{-11}}\right)^{\frac{1}{4}} e^{-\frac{\pi}{4}(\kappa-36)},}
    \label{eq:causality_turning_point}
\end{align}
where we use Eqs.~ \eqref{eq:freq_temp_relation} and  \eqref{eq:metastable_breaking_temperature} and multiply by a factor of 0.01 to account for the effect of loop evolution and the higher-mode summation, such that it fits well the numerical result.
Below $f_{\rm GW}^{\rm brk}$,  the GWB spectrum is not generated by loops and is dominated by the causality tail ($\Omega_{\rm GW} \propto f_{\rm GW}^3$).
Overall, the spectral shape of GWB from metastable local strings follows the asymptotic behavior 
\begin{align}
    \Omega_{\rm GW}(f_{\rm GW}) \propto \begin{cases}
        f_{\rm GW}^3 ~ &{\rm for } ~ f_{\rm GW} \ll f_{\rm GW}^{\rm brk},\\
        f_{\rm GW}^2 ~ &{\rm for } ~ f_{\rm GW}^{\rm brk} \ll f_{\rm GW} \ll f_{\rm GW}^{\rm meta},\\
        {\rm 1} ~ &{\rm for } ~ f_{\rm GW}^{\rm meta} \ll f_{\rm GW} \ll f_{\rm GW}^{\rm UV},\\
        f_{\rm GW}^{-1/3} ~ &{\rm for } ~ f_{\rm GW}^{\rm UV} \ll f_{\rm GW}.
    \end{cases}
\end{align}

In addition to the condition on the rate $\Gamma_d l > H$, the monopole nucleation must be allowed energetically, i.e., the string with length $l$ and energy $\mu l$ can convert into monopole pair of energy $2m_M$ if 
$l > l_{\rm min}^{\rm brk} =  {2 m_M}/{\mu} = ({2}/{m_{\rm Pl}})\sqrt{{\kappa}/{G\mu}}$,
where $m_{\rm Pl} = G^{-1/2}$ is the Planck mass.
A loop formed at Hubble scale $H(t)$ with size $\alpha t \simeq \alpha H^{-1}$ supports the monopole-pair creation if $\alpha t > l_{\rm min}^{\rm brk}$ or equivalently if a loop is formed below temperature
\begin{align}
    T_{\rm pair} \simeq 8.5 \times 10^{15} ~ {\rm GeV} \sqrt{\frac{\alpha}{0.1}\left(\left[\frac{106.75}{g_*(T)}\right]\left[\frac{G \mu}{10^{-6}}\right]\left[\frac{36}{\kappa}\right]\right)^{\frac{1}{2}}},
\end{align}
where the radiation domination is assumed. 
The metastability cutoff saturates at $f_{\rm GW}(T_{\rm pair})$ if $T_{\rm pair} < T_{i,\rm sup}$ in Eq.~\eqref{eq:metastability_GW_suppression_temperature} or equivalently when
\begin{align}
    G\mu \lesssim 9.5 \times 10^{-25} \left(\frac{\alpha}{0.1}\right)^{2}\left(\frac{\kappa}{16}\right)e^{-\pi(\kappa-16)}.
\end{align}
The lower bound on the string length $l_{\rm min}^{\rm brk}$ allowing for monopole-pair creation does not affect the parameter space of the metastable string considered in this paper.

So far, we only consider the GWB produced from loops in the metastable local-string network. In addition, the string segments with monopoles on both ends---produced after time $t_{\rm brk}$---can also lead to GWBs \cite{Leblond:2009fq,Buchmuller:2021mbb}. We provide a detailed discussion regarding the GWB from segments in appendix~\ref{app:segments} and show the comparison between GWBs from loops and GWBs from the two types of segments in Fig.~\ref{fig:local_GW_GWB_segment}. See also Fig.~(5) of \cite{Buchmuller:2021mbb}. Using the simple assumption that the GW-emission from segments is the same as from loops ($\Gamma_{\rm seg} = \Gamma = 50$), the segments's GWB can dominate the IR tail and the flat plateau of loops' contribution for $G\mu \gtrsim 10^{-9}$ and $G\mu \gtrsim 6 \times 10^{-5}$, respectively. If the IR tail of GWB is detected, this additional GWB could make the inferred energy scale $\eta$ overestimated by a factor of $\mathcal{O}(10)$. Similarly, the effect on the metastability cut-off makes the inferred $\kappa$ misidentified by a factor of $\sim 3-6$ for $G\mu \simeq 10^{-5}$. As discussed in appendix~\ref{app:segments}, these results of segments' GWB have to be taken with care because of several uncertainties. The GW emission frequency from segments could be different from that of loops and lead to the frequency shift of the spectrum. The realistic $\Gamma_{\rm seg}$, which can be different from $\Gamma = 50$, modulates the GWB amplitude; see Fig.~\ref{fig:local_GW_GWB_segment}-right. Therefore, this work neglects the GWB from segments and only discusses the conservative detection's prospects and bounds derived from the GWB of loops of metastable local strings.

\subsection{Large GWB in the UHF regime}

\begin{figure*}[t]
\includegraphics[width=\linewidth]{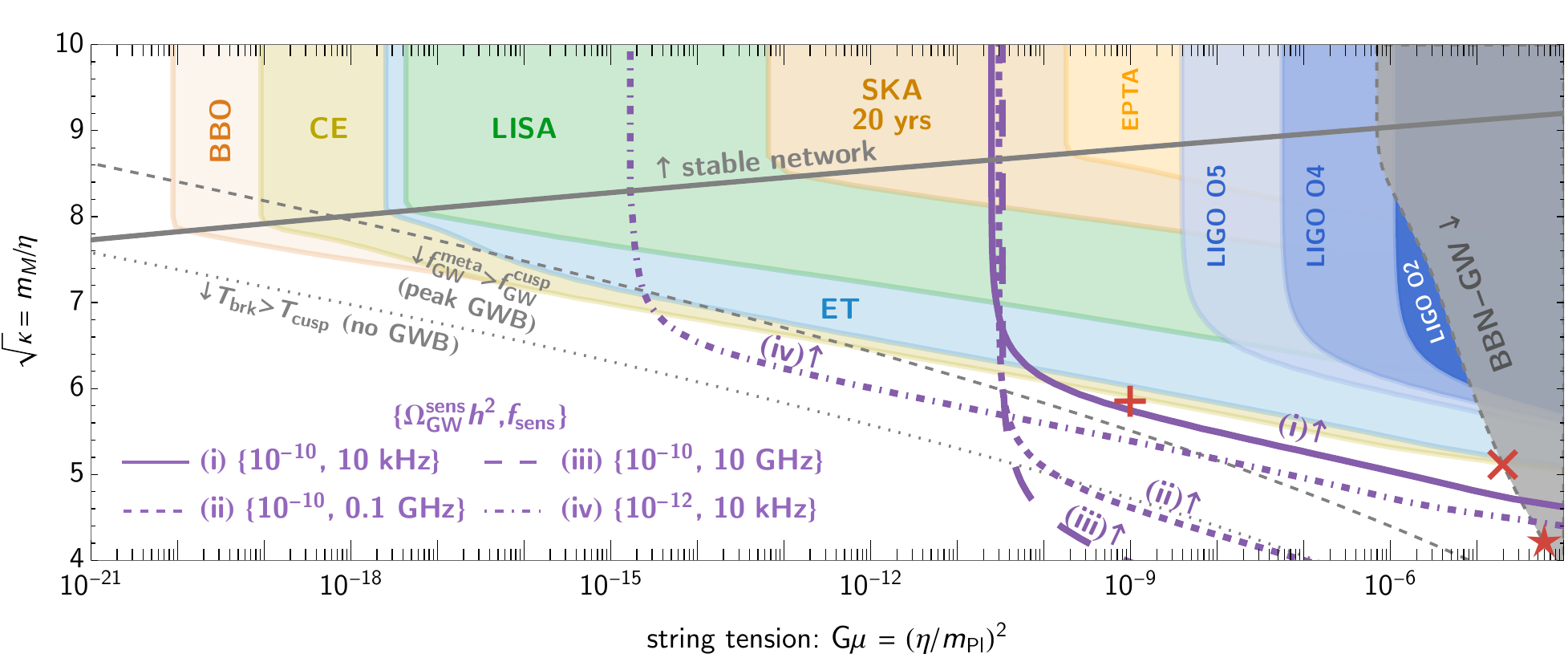}\\[-1em]
\caption{Future detectability regions of metastable local-string GWB by several experiments (only the gray region from the BBN bound and the dark blue from LIGO O2 are excluded). The criterion for detection and constraints is any part of the spectrum lying in the sensitivity and constrained regions. The purple solid, dashed, long-dashed, and dot-dashed lines illustrate the extended detectable region from hypothetical UHF experiments operating at different $\{\Omega_{\rm GW}^{\tt sens}h^2,f_{\tt sens}\}$. The red symbols are associated with the benchmark GW spectra in Fig.~\ref{fig:local_GW_GWB}. 
Assuming loops also produce particles through cusps, the dashed gray line marks the region where $f_{\rm GW}^{\rm brk}>f_{\rm GW}^{\rm cusp}$ and the GW is suppressed according to Eq.~\eqref{eq:GW_suppressed_two_cutoffs} [see also Fig.~\ref{fig:local_GW_GWB_combined_cutoff}].
Below the gray dotted line ($T_{\rm brk} > T_{\rm cusp}$), all loops produced before the network decay release energy via particle production and thus do not generate GW signal.
An earlier version of this plot appeared in \cite{Buchmuller:2019gfy}; we extended it by displaying the gained reach from UHF detectors.}
\label{fig:local_GW_GWB_contour}
\end{figure*}

Three benchmark GWB spectra of metastable strings are shown in Fig.~\ref{fig:local_GW_GWB}, in units of the energy-density fraction $\Omega_{\rm GW}h^2$ as well as of the characteristic strain of GW $h_c$. 
We can see that the IR tails of these spectra follow the scaling $f_{\rm GW}^2$ and $f_{\rm GW}^3$, with the turning-point \eqref{eq:causality_turning_point}, as discussed above.
These benchmark cases evade the current or planned GW experiments, as shown in Fig.~\ref{fig:local_GW_GWB_contour}.
However, the entire parameter space above the kilohertz can be populated by the GWB from metastable strings.

The only upper bound on the GW amplitude comes from the BBN bound in Eq.~\eqref{eq:Neff_bound}. Using that the GWB shape is a flat plateau with the UV formation cutoff $f_{\rm GW}^{\rm form}$ and the IR metastability cutoff $f_{\rm GW}^{\rm meta}$, the ``\textsc{bbn-gw}" becomes 
\begin{align}
    h^2\Omega_{\rm GW} <  \frac{5.6 \times 10^{-6} \Delta N_{\rm eff}}{\log\left[\frac{f_{\rm GW}^{\rm form}}{\max\left(f_{\rm GW}^{\rm BBN},f_{\rm GW}^{\rm meta}\right)}\right]},
    \label{eq:BBN_GW_CS}
\end{align}
where $f_{\rm GW}^{\rm BBN}$ is the GW frequency at BBN scale [using $T_{\rm BBN} \simeq {\rm MeV}$ in Eq.~\eqref{eq:freq_temp_relation}].
For illustrative purposes, we show the BBN bound of $h^2\Omega_{\rm GW}(f_{\rm GW}^{\rm meta})$ in gray regions in our figures.
A UHF GW experiment with sensitivity  a few orders of magnitude below the BBN bound would yield information about  cosmic strings and monopoles in the early Universe.

To demonstrate the discovery potential of UHF-GW experiments, we assume some hypothetical experiments operating in the UHF regime with sensitivity and  central frequency
\begin{align}
    \left\{\Omega_{\rm GW}^{\tt sens}h^2, f_{\tt sens}\right\} ~ ~ {\rm within} ~ ~ \left[\frac{f_{\tt sens}}{2}, \frac{3f_{\tt sens}}{2}\right],
    \label{eq:UHF_experiment}
\end{align}
which are shown as purple rectangles in the GWB spectra plots and purple lines in the parameter space plot of Fig.~\ref{fig:local_GW_GWB_contour}. We assume four benchmark sensitivities curves with $\left\{\Omega_{\rm GW}^{\tt sens}h^2, f_{\tt sens}\right\}$ = (i) $\{10^{-10}, 10 ~ {\rm kHz}\}$, (ii) $\{10^{-10}, 0.1 ~ {\rm GHz}\}$,  (iii) $\{10^{-10}, 10 ~ {\rm GHz}\}$, and (iv) $\{10^{-12}, 10 ~ {\rm kHz}\}$.

\subsection{Peak-shape GWB}
\label{sec:local_string_peak_GWB}
The large GWB spectra allowed by metastability can exhibit an interesting ``peak" shape when including the UV cutoffs of section~\ref{sec:GW_HF_cutoff}.
This happens if the IR cutoff from the metastability is lower than the UV cutoffs, i.e., $f_{\rm GW}^{\rm meta} \leq f_{\rm GW}^{\rm UV}$.
This section focuses on the cusp cutoff~\eqref{eq:freq_cusp}, while this result can be easily extended to other UV cutoffs.

By imposing the cusp cutoff, the GWB from metastability gets suppressed into the $\Omega_{\rm GW} \propto f_{\rm GW}^{-1/3}$ scaling for $f_{\rm GW} > f_{\rm GW}^{\rm cusp}$. 
The GW spectrum peaks at the frequency $f_{\rm GW}^{\rm cusp}$ and has the approximated amplitude 
\begin{align}
    \Omega_{\rm GW}^{\rm peak} \simeq \Omega_{\rm GW}^{\rm stable} \left({f_{\rm GW}^{\rm cusp}}/{f_{\rm GW}^{\rm meta}}\right)^{2},
    \label{eq:GW_suppressed_two_cutoffs}
\end{align}
assuming $f_{\rm GW}^{\rm cusp} > f_{\rm GW}^{\rm brk}$ (i.e., the IR tail retains the $\Omega_{\rm GW} \propto f_{\rm GW}^{2}$ scaling), otherwise one needs to account for the causality tail, discussed around Eq.~\eqref{eq:causality_turning_point}.
Using Eqs.~\eqref{eq:freq_cusp} and \eqref{eq:metastable_cutoff_freq}, we have the suppression factor as $(f_{\rm GW}^{\rm cusp}/f_{\rm GW}^{\rm meta})^2 \propto (G\mu)^{5/2}\exp(\pi \kappa/2)$. For a fixed $\kappa$, the peak-shape GWB is present for
\begin{align}
    \boxed{G\mu \lesssim 2.8 \times 10^{-11} \, \beta_c^{2/5} e^{-\frac{\pi}{5}(\kappa - 36)} ~ ~ {\rm (suppressed)}.}
    \label{eq:peak_GW_spectrum_Gmu_sup}
\end{align}
Furthermore, all loops decay into  particles rather than  GW  if $T_{\rm cusp} < T_{\rm brk}$, i.e.,
\begin{align}
    G\mu \lesssim 2.1 \times 10^{-14} \, \sqrt{\beta_c} e^{-\frac{\pi}{4}(\kappa - 36)} ~ ~ {\rm (no ~ GWB)},
\end{align}
where we use Eqs.~\eqref{eq:temperature_cusp} and \eqref{eq:metastable_breaking_temperature}.
Fig.~\ref{fig:local_GW_GWB_combined_cutoff} shows examples of peak GWB spectra calculated numerically for a fixed $G\mu$, a similar plot with a fixed metastability parameter $\kappa$ is shown in Fig.~\ref{fig:local_GW_GWB_combined_cutoff_app} of appendix~\ref{app:peak_GWB}.
In Fig.~\ref{fig:local_GW_GWB_contour}, we indicate below the gray dashed line the region of parameter space where, due to the cusp cutoff, the GWB spectra are peak-shaped ($f_{\rm GW}^{\rm meta} > f_{\rm GW}^{\rm cusp}$), while below the dotted line, there is no GWB, corresponding to  ($T_{\rm brk} > T_{\rm cusp}$).

As shown in Figs.~\ref{fig:local_GW_GWB} and \ref{fig:local_GW_GWB_contour}, the GWB from metastable local strings can have a large amplitude in the UHF regime, saturating the BBN bound when the metastability parameter $\kappa$ is small enough. For example, the optimal benchmark point ``$\times$" has $G\mu \simeq 10^{-5}$, corresponds to the GUT symmetry-breaking scale $\eta \sim 10^{16}~ {\rm GeV}$, while the string formation scale is close to the monopole's scale. Detecting any UV cutoff on the GWB spectrum will yield information about the nature of cosmic strings: the underlying field theory and its interactions. We will now investigate the possibility of reconstructing the scalar potential from the detectable UV cutoffs---using the UHF-GW experiments.

\begin{figure}[t]
\includegraphics[width=\linewidth]{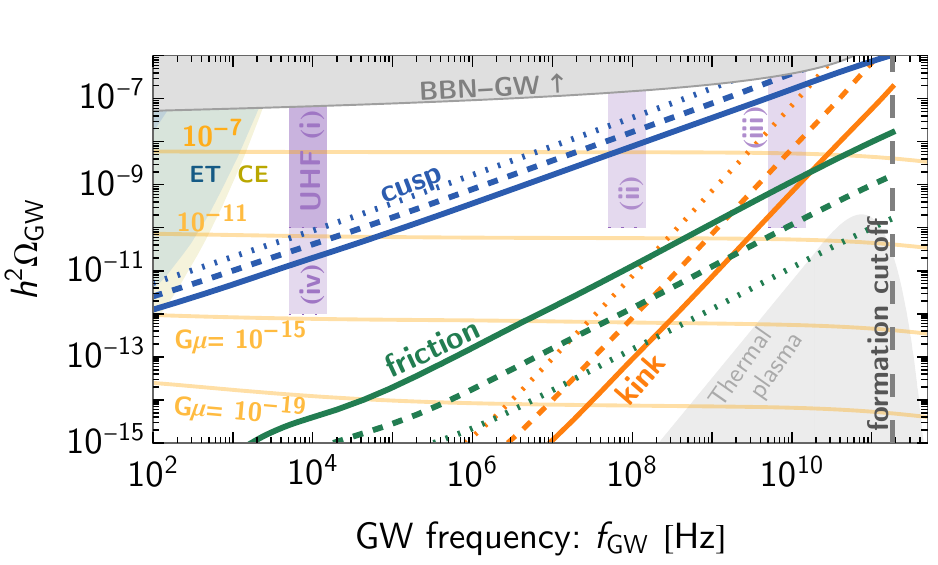}\\[-1em]
\caption{By varying the string tension $G\mu$, positions of the cutoffs---cusp, kink, and friction on the local-string GWB [see Eqs.~\eqref{eq:kink_cutoff_position}, \eqref{eq:cusp_cutoff_position}, and \eqref{eq:friction_cutoff_position}]. For cusp and kink, the solid, dashed, and dotted lines correspond respectively to $\beta_c,\beta_k$ = 1,10,100. The friction cutoff with $\beta_{\rm fric} = 0.01,0.1,1$ is shown as solid, dashed, and dotted lines, respectively. The yellow spectra are those of the stable-string network with the formation cutoff.}
\label{fig:local_GW_UV_cutoff}
\end{figure}

\section{Reconstructing the scalar potential of cosmic strings}
\label{sec:recontruct}
Given the discussion of the previous section, it is clear that the position of the UV cutoff depends on the parameters of the microscopic theory. 
Fig.~\ref{fig:local_GW_UV_cutoff} show the locations of the UV cutoffs when varying $G\mu$ from cusps, kinks,  friction, and formation [Eqs.~\eqref{eq:formation_cut_off}, \eqref{eq:kink_cutoff_position}, \eqref{eq:cusp_cutoff_position}, and \eqref{eq:friction_cutoff_position}] in the $\{f_{\rm GW}, \Omega_{\rm GW}h^2\}$ plane.
For each cutoff, we also show how the coefficients $\beta_k$, $\beta_c$, and $\beta_{\rm fric}$ change the positions of the cutoffs.
The GWBs from stable strings are in yellow lines as references.
For kink and cusp cutoffs, we vary $\beta_c$ and $\beta_k$ to values larger than 1 as they grow with a larger number of kinks and cusps and with smaller self-coupling $\lambda$, as shown in Eq.~\eqref{eq:beta_cusp_kink}.
We vary $\beta_{\rm fric}$, which is proportional to the interaction cross-section, from $10^{-3}$ to $10^{3}$.

By locating the cutoff, it is conceivable to infer from these measurements [using Eq.~\eqref{eq:beta_cusp_kink}], the values of parameters, such as the quartic coupling $\lambda$ of the scalar field as well as the vacuum expectation value $\eta$ which gives the scale of $U(1)$ symmetry breaking.

\begin{figure*}[t!]
	\centering
 \includegraphics[width=\linewidth]{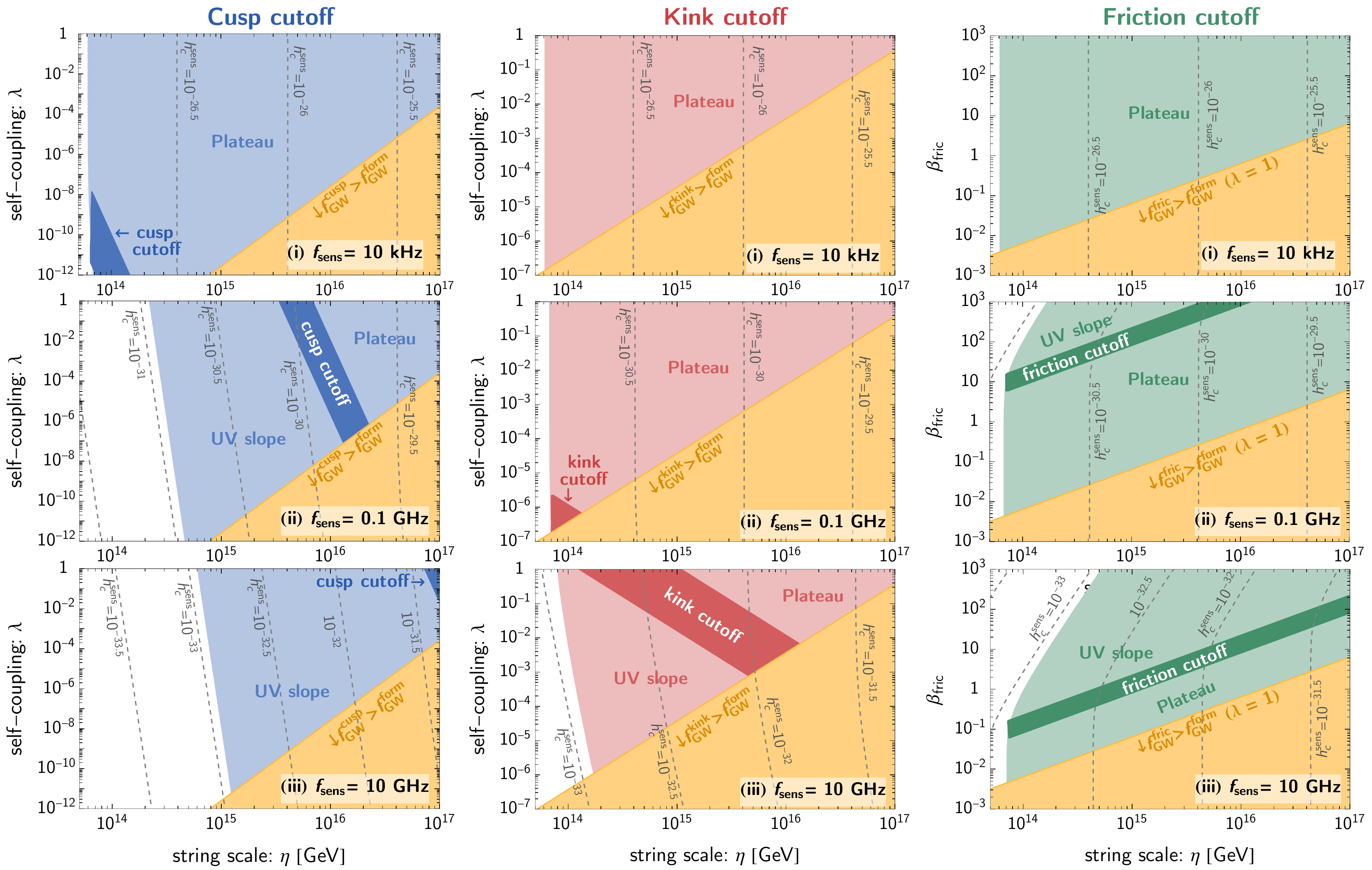}\\[-0.5em]
\caption{Colored region in each plot shows the detectable local-string GWB spectrum, including cusp (left-column), kink (middle-column), and friction (right-column) cutoffs. 
The detectable region splits into three parts depending on the feature of the GWB spectrum: the plateau (flat slope $\Omega_{\rm GW} \propto f_{\rm GW}^{0}$), the cutoff (where the slope changes), and the UV slope ($\Omega_{\rm GW} \propto  f_{\rm GW}^{-1/3}$). 
We assume the hypothetical UHF experiments operating at $\Omega_{\rm GW}^{\tt sens} = 10^{-10}$ with central frequencies: (i) 10 kHz, (ii) 0.1 GHz, and (iii) 10 GHz. The yellow region shows where these UV cutoffs are higher than the formation cutoff. The gray contours are the characteristic strain $h_c$ that would be detected in such experiments.}
    \label{fig:recontr}
\end{figure*}

The possible strategy to pindown the cutoff's position is to measure the GW spectrum at two different frequencies.
If one detector can observe the flat part while the other observes the UV slope, we can deduce $G\mu$ (or $\eta$) and  $\beta$ (once $G\mu$ is known), respectively. 
Fig.~\ref{fig:recontr} shows the regions of the $\eta, \lambda, \beta_{\rm fric}$ parameter spaces that can be probed by the hypothetical UHF-GW experiments operating with $\Omega_{\rm GW}^{\tt sens} = 10^{-10}$ and $f_{\rm GW}^{\tt sens} =$ (i) 10 kHz, (ii) 0.1 GHz, and (iii) 10 GHz. The gray dashed contours correspond to the characteristic strain $h_c^{\tt sens}$ of the detected signal.
As an example, let us assume that the detector (i) sees a flat GWB with $h_c = 10^{-26}$; that is $\eta \simeq 4 \times 10^{15} ~ {\rm GeV}$. 
Moreover, if the detector (iii) observes GWB with $h_c = 10^{-32.5}$, we can infer that the cosmic strings have cuspy loops and the underlying scalar potential has $\eta=10^{15}$ GeV with $\lambda \simeq 10^{-8}$. From this particular observation, we would conclude the absence of a kink or a friction cutoff.

\section{Metastable global-string GWB: heavy axions}
\label{sec:global_metastable_network}
We now move to the case where the $U(1)$ symmetry is global; we can then identify the symmetry breaking scale $\eta$ to the axion decay constant $f_a$.
When it breaks spontaneously after inflation at energy scale $f_a$ (so-called  ``post-inflationary" axion scenario), the network of global cosmic strings also forms loops and generates a GWB. 
The main difference with the local case is that the angular mode of the complex scalar field $\Phi$ associated with cosmic strings, the axion, is a Nambu-Golstone boson \cite{Peccei:1977hh, Peccei:1977ur, Weinberg:1977ma, Wilczek:1977pj}. It receives mass $m_a$ from axion shift-symmetry breaking dynamics and generates domain walls (DWs) \cite{Kibble:1976sj,Kibble:1980mv,Vilenkin:1982ks}. After formation, DWs attach to cosmic strings and make them collapse entirely when $H \sim m_a$ \cite{Vilenkin:1982ks,Hiramatsu:2012gg} or equivalently at temperature
\begin{align}
    T_{\rm dec} \simeq 9 \times 10^8 ~ {\rm GeV} \left[\frac{106.75}{g_*(T_{\rm dec})}\right]^{\frac{1}{4}}\left(\frac{m_a}{\rm GeV}\right)^{\frac{1}{2}},
    \label{eq:Tdec_ma}
\end{align}
if the domain-wall number is unity\footnote{For the domain-wall number greater than unity, the string-wall system is stable, and the dominant GWB comes from domain walls \cite{Hiramatsu:2010yz,Kawasaki:2011vv,Hiramatsu:2012sc,Hiramatsu:2013qaa}. We will not consider such a scenario in this work.}.

The metastability of the cosmic-string network is therefore built-in automatically  and can be included in the GWB calculation by introducing the cutoff in the loop number density after time $t_{\rm dec} \equiv t(T_{\rm dec})$
\begin{align}
    n_{\rm loop}^{\rm dec} = n_{\rm loop}^{\rm stable} \Theta(t_{\rm dec} - t_i),
    \label{eq:global_loop_density}
\end{align}
where $t_i$ is the time when loops are produced, and $n_{\rm loop}^{\rm stable}$ is the loop number density of the stable network from the VOS model, calculated as in \cite{Gouttenoire:2019kij}.
Note that Eq.~\eqref{eq:global_loop_density} is similar to the case of local strings in Eq.~\eqref{eq:local_loop_density}, except the factor $\mathcal{E}$ which suppresses loop density over time.
The global-string loops decay quickly into GW and goldstone particles, such that the GW is determined by the number density of loops at the time of production $t_i$.

With no loop production below $T_{\rm dec}$, the GWB gets cut into the causality tail ($\Omega_{\rm GW} \propto f_{\rm GW}^3$) at frequency lower than 
\begin{align}
    f_{\rm GW}^{\rm RD}(T_{\rm dec}) \simeq 9.4 ~ {\rm kHz} \left(\frac{\alpha}{0.1}\right)\left(\frac{m_a}{\rm GeV}\right)^{\frac{1}{2}},
\label{eq:freq_cs_axion_mass}
\end{align}
using $T_{\rm dec}$ in Eq.~\eqref{eq:freq_temp_relation} where the superscript ``\textsc{rd}" reminds us that it assumes the radiation-dominated Universe at high energies as in the standard cosmological model. This is not always the case for the heavy axion, where the frequency can be further shifted by the axion matter-domination era; see below.
The asymptotic behavior of the spectral shape is thus
\begin{align}
    \Omega_{\rm GW}(f_{\rm GW}) \propto \begin{cases}
        f_{\rm GW}^{3} ~ &{\rm for } ~ f_{\rm GW} \ll f_{\rm GW}^{\rm dec},\\
        \mathcal{D}^3 ~ &{\rm for } ~ f_{\rm GW} \gg f_{\rm GW}^{\rm dec},
    \end{cases}
\end{align}
where $\mathcal{D}$ is the log-dependent factor defined in Eq.~\eqref{eq:log_dependence_D_factor}.
That is, the GWB from metastable global strings exhibits a peak-shape spectrum, with peak frequency $f_{\rm GW}^{\rm RD}$ and peak amplitude $\Omega_{\rm GW}^{\rm RD, peak}$ estimated by Eq.~\eqref{eq:GWB_stable}. In this work, we determine the peak amplitude from the numerically generated GWB.

From Eq.~\eqref{eq:freq_cs_axion_mass}, we find that 
the GWB from global axionic strings with axion mass $m_a \gtrsim 1 ~ {\rm GeV}$  will not appear in  GW experiments ($f_{\rm GW} \gtrsim 9 ~ {\rm kHz}$) and could in principle be a well-motivated target for UHF-GW detectors.
However, the decay of the string network produces heavy axions, which behave as non-relativistic matter and lead to an axion matter-domination (MD) era.
We will discuss first the axion-MD era from the string-network decay and later the shifted and  suppressed\footnote{The axion MD also leads to the modified causality tail \cite{Hook:2020phx, Racco:2022bwj, Franciolini:2023wjm}, although this feature's signal is even more suppressed. For completeness, we discuss it in appendix~\ref{app:axion_MD_IR_tail} where its estimated position in the GWB spectrum is shown in Fig.~\ref{fig:axion_MD_IR_tail}.} GWB.

\begin{figure}[b!]
\includegraphics[width=\linewidth]{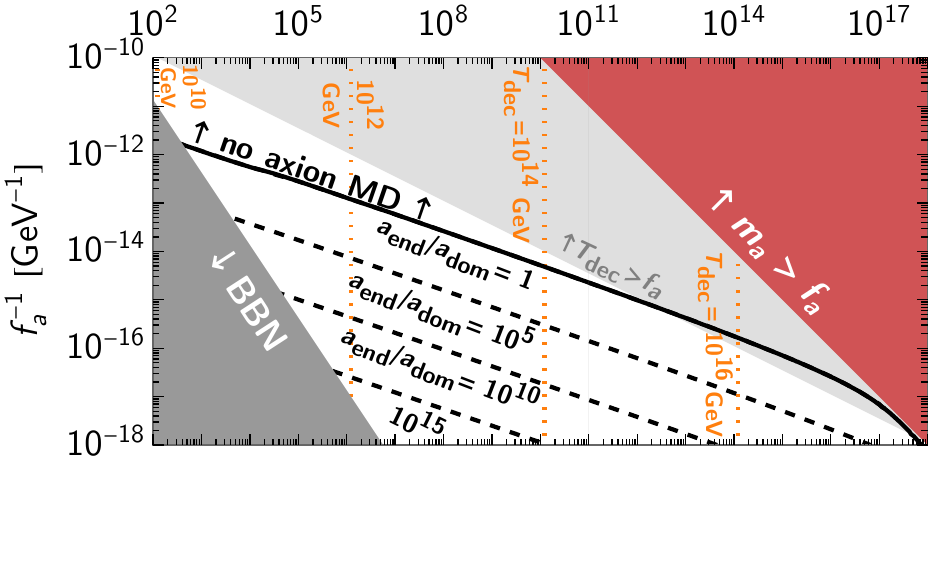}\\[-4em]
\includegraphics[width=\linewidth]{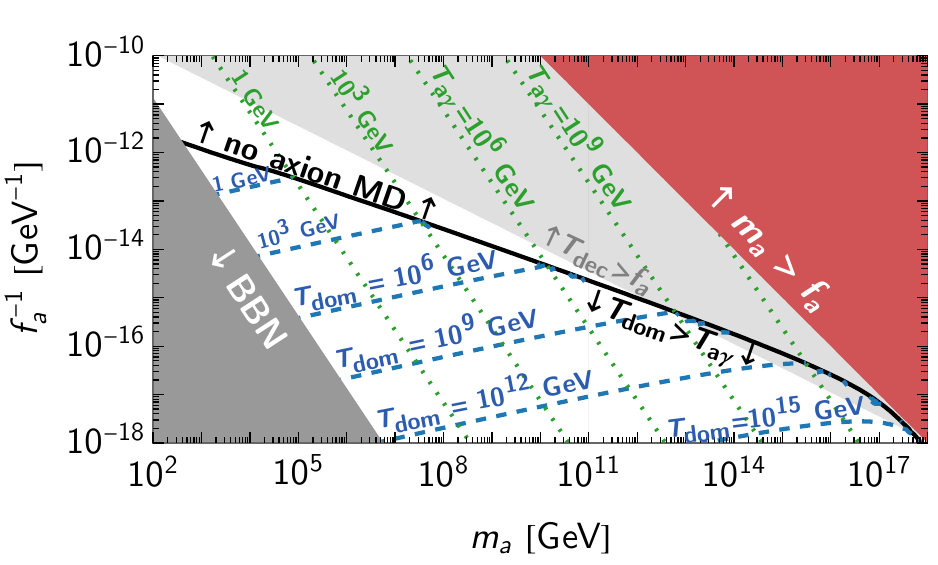}\\[-0.5em]
\caption{Parameter space of the heavy axion below the solid black line gives rise to the axion MD era with duration $a_{\rm end}/a_{\rm dom}$ [marked by the dashed black line], which decays before the BBN (otherwise, it is constrained as in the dark gray region.)
The axions are produced from string-network decay as a subdominant energy density at $T_{\rm dec}$ (vertical dotted orange lines in the top panel). When they dominate the Universe, the axion MD era starts at $T_{\rm dom}$ (blue dashed lines) and later ends when they decay at $T_{a \gamma}$ (green dotted lines).
The lighter gray region has the string network decaying before the string formation at temperature $f_a$. The perturbativity of the axion effective-field-theory description breaks down in the red region.}
\label{fig:axion_MD}
\end{figure}

\subsection{Heavy axion matter-domination era}
The string-wall network at $T_{\rm dec}$ having energy density $\rho_{\rm net}(T_{\rm dec}) \simeq \mu(T_{\rm dec})/t_{\rm dec}^2 \simeq G\mu(T_{\rm dec}) \rho_{\rm tot}(T_{\rm dec})$ decays into non-relativistic axions (each of energy $\sim H \sim m_a$ \cite{Gelmini:2022nim, Davis:1986xc, Yamaguchi:1998gx, Hiramatsu:2010yu}). Its energy density $\rho_{a}(T_{\rm dec}) \simeq \rho_{\rm net}(T_{\rm dec})$ starts to redshift as $\rho_a \propto a^{-3}$ and dominates the thermal plasma energy density at temperature
\begin{align}
    T_{\rm dom} \simeq T_{\rm dec} G\mu(T_{\rm dec}) \left[\frac{g_*(T_{\rm dec})g_{*s}(T_{\rm dom})}{g_*(T_{\rm dom})g_{*s}(T_{\rm dec})}\right],
    \label{eq:T_dom_axion_MD}
\end{align}
leading to the axion MD era.

The axion MD era lasts until the axions decay, e.g., via $\mathcal{L}\supset - g\theta F\tilde{F}/4$ into SM photons which happens when the corresponding decay rate---$\Gamma_{a\gamma} = m_a^3 g^2/(64\pi)$ \cite{Cadamuro:2011fd} with $g_{a\gamma} = {1.92 \alpha_{\rm em}}/({2 \pi f_a})$---is comparable to the Hubble rate. I.e., $\Gamma_{a \gamma} \simeq H_{\rm end}$. Using that $3 M_{\rm Pl}^2 H_{\rm end}^2 = \frac{\pi^2}{30}g_*(T_{a\gamma})T_{a\gamma}^4$, we obtain the temperature of the thermal plasma right after axions' decay as
\begin{align}
    T_{a\gamma} \simeq 4.2 ~ {\rm MeV} \left[\frac{106.75}{g_*(T_{a\gamma})}\right]^{\frac{1}{4}} \left(\frac{m_a}{\rm TeV}\right)^{\frac{3}{2}}\left[\frac{10^{12 }~ \rm GeV}{f_a}\right].
\end{align}
The axion MD exists  if $T_{\rm dom} > T_{a\gamma}$ with the inverse duration, defined by 
\begin{align}
    \boxed{\mathcal{B} \equiv \frac{a_{\rm dom}}{a_{\rm end}} = \left[\left(\frac{3\sqrt{10}}{64 \pi^2}\right)\frac{m_a^3 g_{a\gamma}^2 M_{\rm Pl}}{g_{*}^{1/2}(T_{\rm dom}) T_{\rm dom}^2}\right]^{\frac{2}{3}} \leq 1,}
    \label{eq:axion_MD_duration}
\end{align}
 which depends on the axion parameters as $\mathcal{B} \propto m_a^{4/3} f_a^{-4}$ [using Eqs.~\eqref{eq:Tdec_ma}, \eqref{eq:T_dom_axion_MD}], up to the log-factor.
Fig.~\ref{fig:axion_MD} shows the heavy-axion parameter space $\{f_a,m_a\}$ giving rise to the axion MD era that ends before BBN.

\begin{figure*}[t!]
	\centering
 {\bf Axion-string GWB}\\[-0.5em]
	\includegraphics[width=0.495\linewidth]{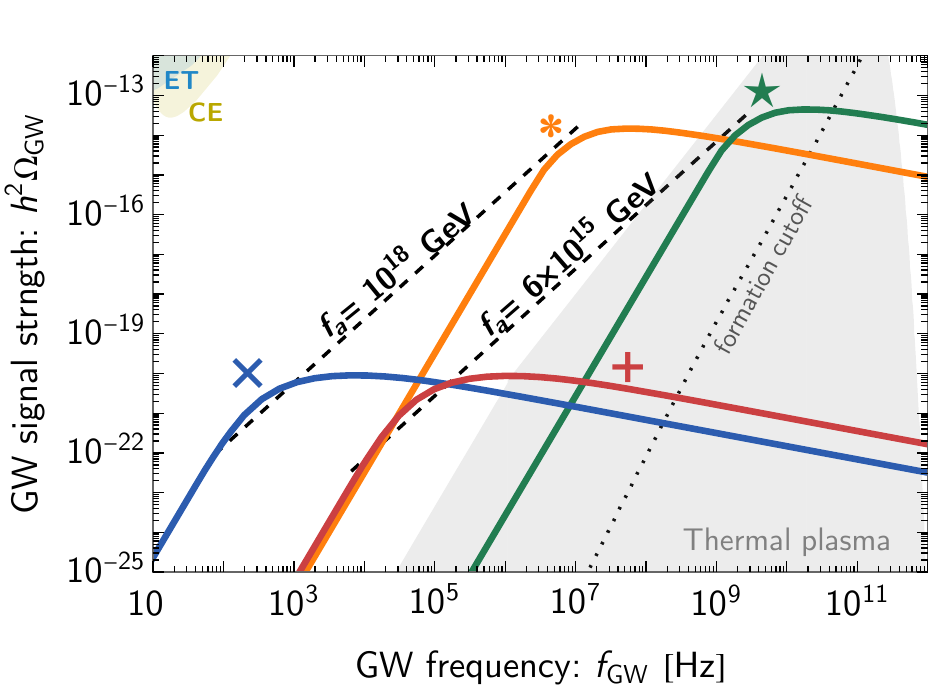}\hfill
 \includegraphics[width=0.495\linewidth]{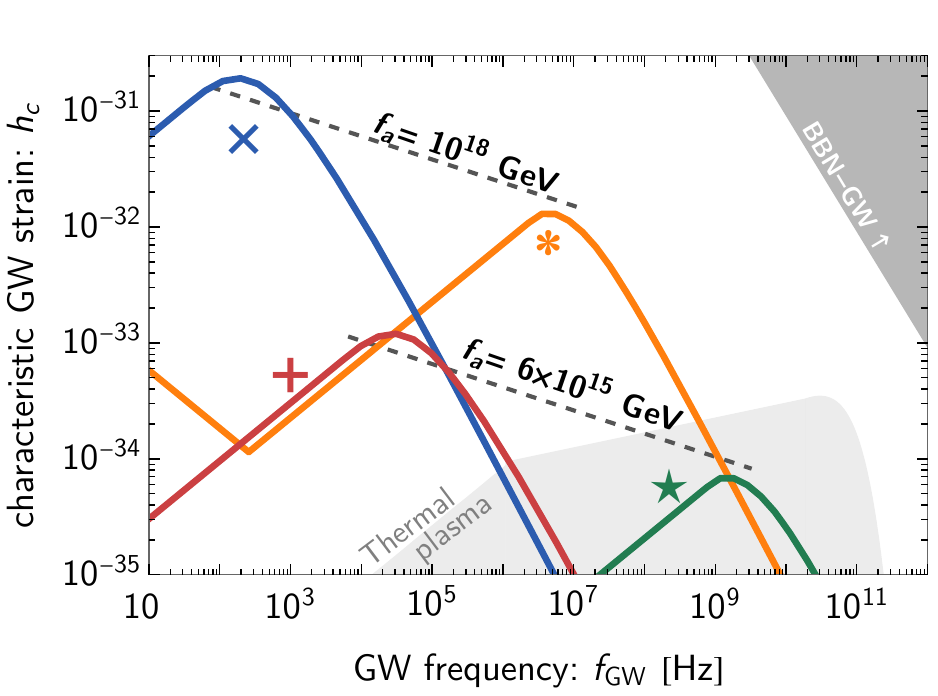}\\[-0.75em]
 \includegraphics[width=0.495\linewidth]{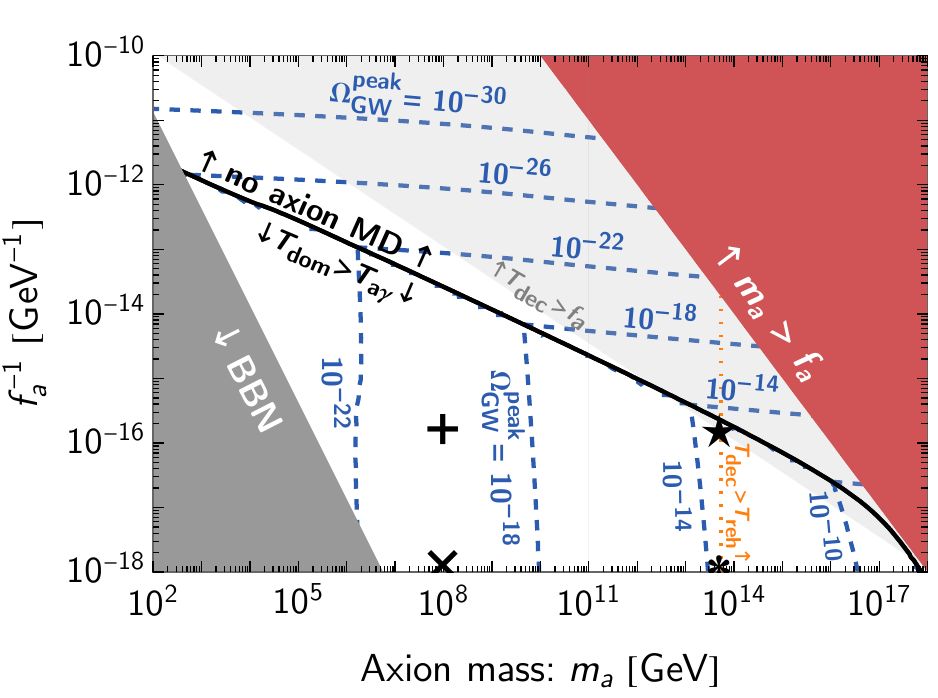}\hfill
\includegraphics[width=0.495\linewidth]{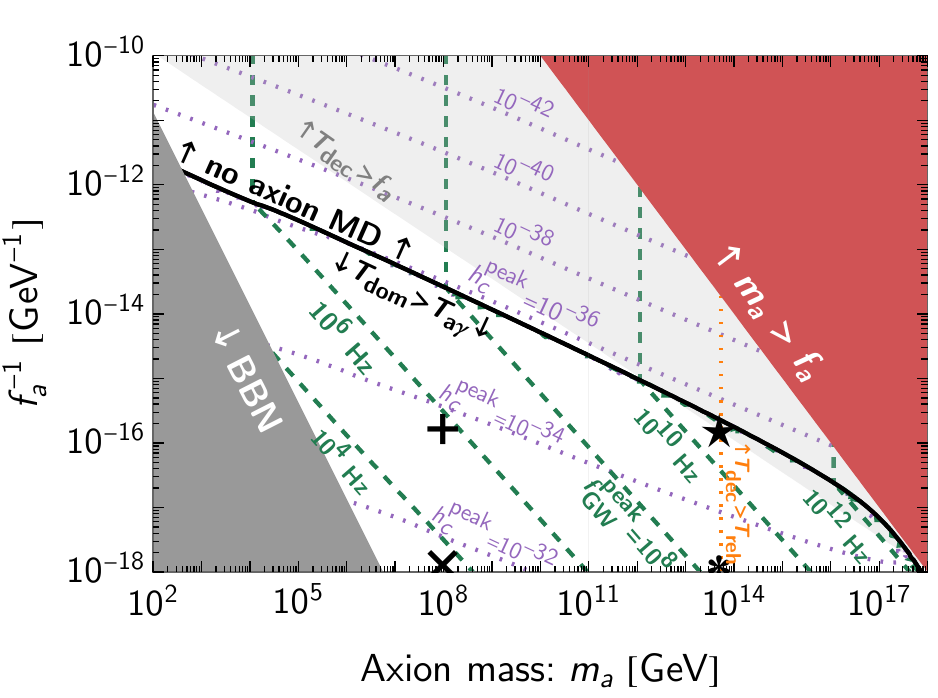}\\[-1em]
\caption{Top panel:  GWB spectra in the energy-density fraction $\Omega_{\rm GW}h^2$ (left) and in the characteristic strain $h_c$ (right) corresponding to the benchmark $\{m_a,f_a\}$ values in Fig.~\ref{fig:global_spectrum_gw}-bottom. The black dotted lines show the peak position of GWB spectrum for a constant $f_a$ while varying $m_a$ within the range where $m_a$ is allowed by $T_{a\gamma}$ above BBN and $T_{\rm dec} < T_{\rm reh}^{\rm max}$; see the bottom panel. The scaling of these lines follow $\Omega_{\rm GW}^{\rm peak} \propto (f_{\rm GW}^{\rm peak})^{8/5}$ and $h_c^{\rm peak} \propto (f_{\rm GW}^{\rm peak})^{-1/5}$ [see main text for derivation]. 
Bottom panel: Contours of the peak amplitude (left) and peak frequency (right) of the GW spectra, i.e., at $f_{\rm GW}(T_{\rm dec})$. The GW from heavy-axion strings cannot generate an observable GW signal even for large $f_a$ due to the heavy dilution from the axion matter-dominated era. ($\star,*$ for $m_a = 5 \cdot 10^{13}$ GeV, $+,\times$ for $m_a =  10^{8}$ GeV)}
    \label{fig:global_spectrum_gw}
\end{figure*}

\subsection{Strongly suppressed GW spectrum}
The GWB from axion strings is produced until the string-network decay. The subsequent axion MD era dilutes both the amplitudes and frequencies of GWB because the Universe undergoes a longer expansion.
To derive the modified GWB, let us consider the GW signals produced at temperature $T_{\rm emit}$ and frequency $f_{\rm GW}^{\rm emit}$.
The frequency of the GW signal today is $f_{\rm GW} = f_{\rm GW}^{\rm emit} (a_{\rm emit}/a_0)$; thus, we can relate the GW frequency in the presence of the axion MD to the one assuming the standard cosmological scenario $f_{\rm GW}^{\rm RD}$ in Eq.~\eqref{eq:freq_cs_axion_mass} as
\begin{align}
    \boxed{f_{\rm GW} = f_{\rm GW}^{\rm RD}  \left[\frac{\mathcal{G}(T_{\rm end})}{\mathcal{G}(T_{\rm dom})}\right]^{\frac{1}{4}}  \mathcal{B}^{\frac{1}{4}}.}
    \label{eq:axion_redshift_frequency_GW}
\end{align}
The energy density of GW signal redshifts as radiation $\Omega_{\rm GW} = ({\rho_{\rm GW}^{\rm emit}}/{\rho_{\rm tot,0}}) ({a_{\rm emit}}/{a_0})^4$. So, the presence of the axion MD era dilutes the GW signal by
\begin{align}
    \boxed{\Omega_{\rm GW}(f_{\rm GW}) = \Omega_{\rm GW}^{\rm RD}[f_{\rm GW}^{\rm RD}(f_{\rm GW})]   \frac{\mathcal{G}(T_{\rm end})}{\mathcal{G}(T_{\rm dom})} \mathcal{B}.}
    \label{eq:axion_redshift_amplitude_GW}
\end{align}
The GWB spectrum at frequencies higher than the IR cutoff gets diluted as a whole and retains its shape. However, below the cutoff, the causality tail of the GWB can also get modified at the frequencies corresponding to the horizon scale during the axion MD era, i.e., the scaling changes from $\Omega_{\rm GW} \propto f_{\rm GW}^3$ during radiation era to $\Omega_{\rm GW} \propto f_{\rm GW}$ during the matter era \cite{Hook:2020phx, Racco:2022bwj}.
We find that the modified causality tail due to the axion MD era appears at extremely low amplitude, e.g., the  change of slope at low $f_{\rm Gw}$ and small $h_c$ in the $*$ spectrum in Fig.~\ref{fig:global_spectrum_gw}-top-right. See more details about the modified causality tail in appendix~\ref{app:axion_MD_IR_tail} and Fig.~\ref{fig:axion_MD_IR_tail}. 

\begin{figure*}[t!]
	\centering
 \includegraphics[width=\linewidth]{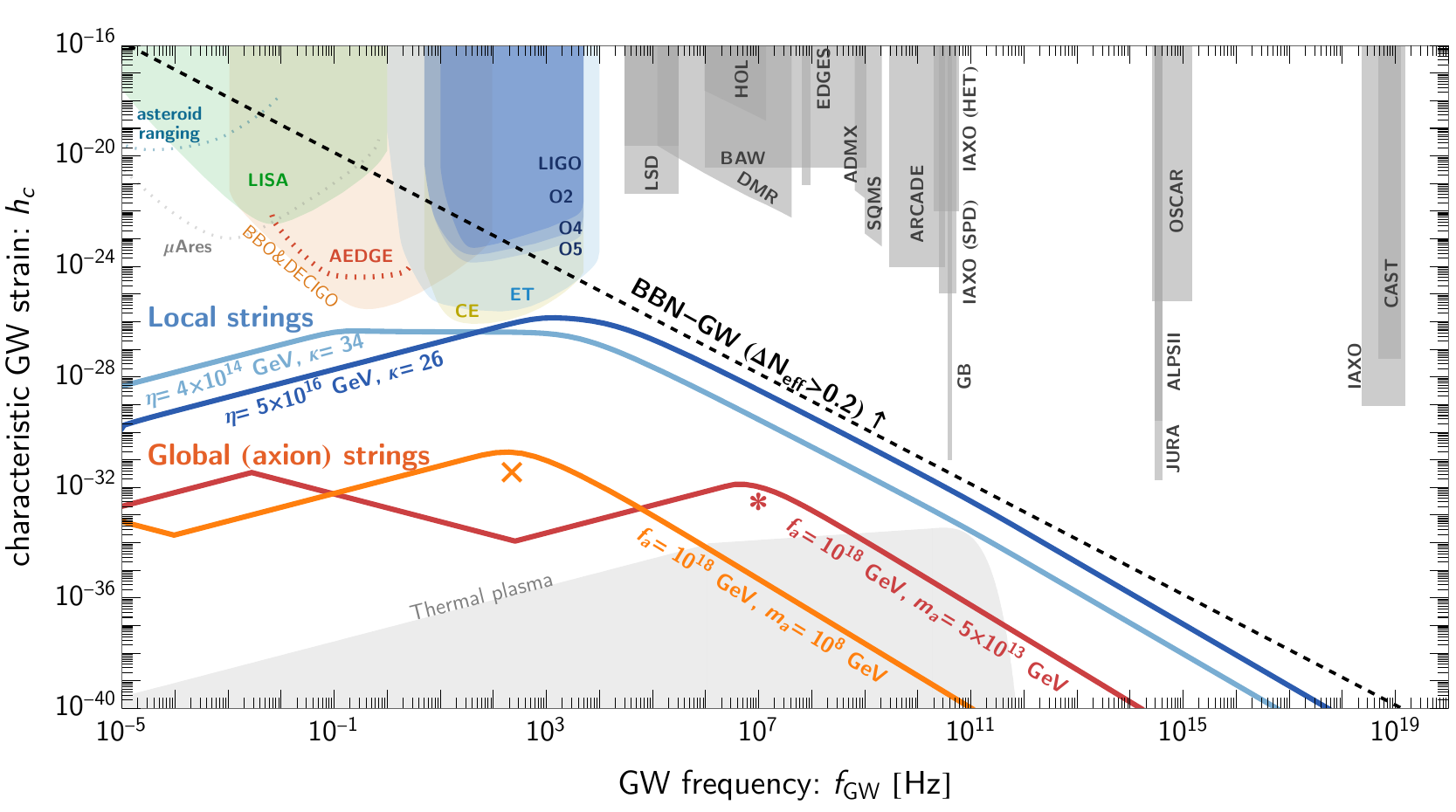}\\[-1em]
\caption{Maximal GWBs expected from cosmic strings (local in blue, global in red/orange) that can arise at ultra-high frequencies without leading to any observable signal in the frequency range of existing or planned interferometers. The spectra from global axionic strings can be large and detectable at lower frequencies for light axions \cite{ Ramberg:2019dgi,Chang:2019mza,Gouttenoire:2019kij,Gelmini:2021yzu,Chang:2021afa,Gorghetto:2021fsn,Gelmini:2022nim,Servant:2023mwt}. However, when requiring a signal that arises only beyond the kHz, this fixes the axion mass to be above the GeV scale to avoid constraints from late decays after BBN. Such heavy axions induce an early temporary matter-domination  era that suppresses the GWB.}
    \label{fig:money_plots}
\end{figure*}

Fig.~\ref{fig:global_spectrum_gw}-top shows the GWB spectra from the heavy-axion strings, corresponding to the benchmark points in Fig.~\ref{fig:global_spectrum_gw}-bottom and accounting already for the dilution from the axion-MD era. 
The gray dotted lines in Fig.~\ref{fig:global_spectrum_gw}-top estimate the peak position of GW spectrum of a constant $f_a$ [$\Omega_{\rm GW}^{\rm peak} \propto (f_{\rm GW}^{\rm peak})^{8/5}$ and $h_c^{\rm peak} \propto (f_{\rm GW}^{\rm peak})^{-1/5}$] deriving from Eq.~\eqref{eq:axion_MD_duration} [$\mathcal{B} \propto m_a^{4/3}$], Eq.~\eqref{eq:freq_cs_axion_mass} [$f_{\rm GW}^{\rm RD} \propto m_a^{1/2}$], and Eq.~\eqref{eq:axion_redshift_frequency_GW}.
From Fig.~\ref{fig:global_spectrum_gw}-bottom, we see that the behavior of GW amplitude when there is no axion [i.e., $\Omega_{\rm GW} \propto f_a^4$ with a mild dependence on $m_a$ in Eq.~\eqref{eq:GWB_stable}] changes to $f_a$-insensitive due to the axion MD era.
Because the dilution factor $\mathcal{B}\propto f_a^{-4}$ ($T_{\rm dom} \propto \mu \propto f_a^2$) cancels the $f_a^4$-dependence of $\Omega_{\rm GW}^{\rm RD}$.

Limiting the string-decay temperature to the maximum reheating temperature (orange dotted line in Fig.~\ref{fig:global_spectrum_gw}-bottom), the benchmark  `$\star$' with $\{m_a, f_a\} = \{5 \cdot 10^{13} ~{\rm GeV}, 4 \cdot 10^{15} ~ {\rm GeV}\}$ gives the largest signal $\Omega_{\rm GW}h^2 \simeq 10^{-14}$. Nonetheless, it is below the thermal plasma GWB (with $T_{\rm reh}^{\rm max} \simeq 6.6\times 10^{15} ~ {\rm GeV}$). On the other hand, the signal with maximum characteristic strain $h_c \simeq 10^{-31}$ is given by the axion mass of $m_a \simeq 10^{8}$ GeV and $f_a \simeq 10^{18}$ GeV. The associated  benchmark spectrum is denoted `$\times$' and has the peak frequency $\gtrsim$ kHz. 
In any case, the axion MD era suppresses the GWB from heavy-axion strings and renders its observability challenging.

\section{Conclusion}
\label{sec:conclude}

The strongest current constraint on UHF GW comes from the $N_{\rm eff}$ bound \eqref{eq:Neff_bound}. Future missions such as CMB stage-4 will strengthen this bound. To get further insight 
on primordial GW would require experiments able to measure part of the spectral shape that encodes information on ultra high energy processes.
 We have shown that that there is a compelling science case for UHF-GW experiments operating with sensitivity slightly below the BBN bound.
 Grand Unification physics in the context of two-step symmetry-breaking events in the $10^{14}-10^{17} $ GeV energy range could lead to large signals close to the BBN bound.
 The maximal signals that can be expected from respectively local and global strings are plotted in the summary figure~\ref{fig:money_plots}, which can be compared to Fig.~\ref{fig:money_plot_intro} that displays other potential primordial sources of GWs at ultra high-frequencies. 
 
 This paper found that the GWB from global axionic strings is limited by the early-matter era induced by heavy axions.
 Local cosmic strings on the other hand appear to be the most promising targets. From the GW amplitude, one can infer the scale of symmetry breaking, while the measurement of the UV cutoff of the GWB could provide microscopic information on the scalar-field couplings.
 Lastly, for such cosmic strings formed at high energy scales, the GWB from string-monopole segments (see appendix~\ref{app:segments}) could provide an additional contribution to that of metastable local-string loops, discussed in the main text. Although segments' GWB gives rise to additional smoking-gun signatures that can be searched for in the future, the GWB calculations are subjected to some theoretical uncertainties that would require further investigation.

\section*{Acknowledgements}
We thank Valerie Domcke, Daniel Figueroa, Camilo Garcia-Cely, and Marco Gorghetto for discussions.
P.S. is supported by Generalitat Valenciana Grant No. PROMETEO/2021/083 and Helmholtz-Promotionspreis 2022. This work is funded by the Deutsche Forschungsgemeinschaft under Germany’s Excellence Strategy---EXC 2121 ``Quantum Universe"---390833306. We appreciate the hospitality of  CERN Theory Department during the ``Ultra-high frequency gravitational waves: where to next?" workshop, when this work was completed.

\appendix

\section{Effect of higher-mode summation}
\label{app:kmax_effect}

\begin{figure}[t!]
\includegraphics[width=0.95\linewidth]{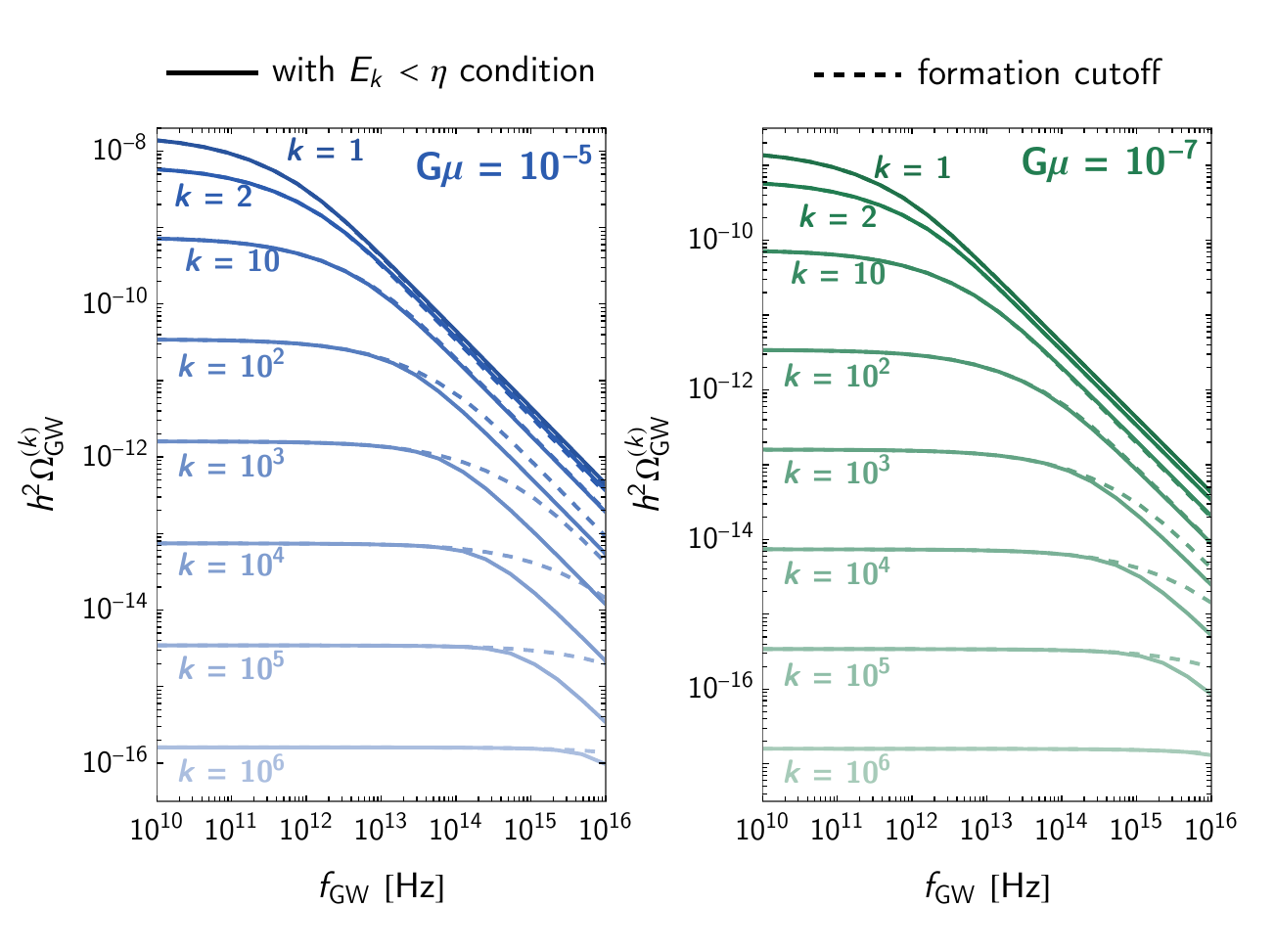}\\[-1.25em]
\includegraphics[width=0.9\linewidth]{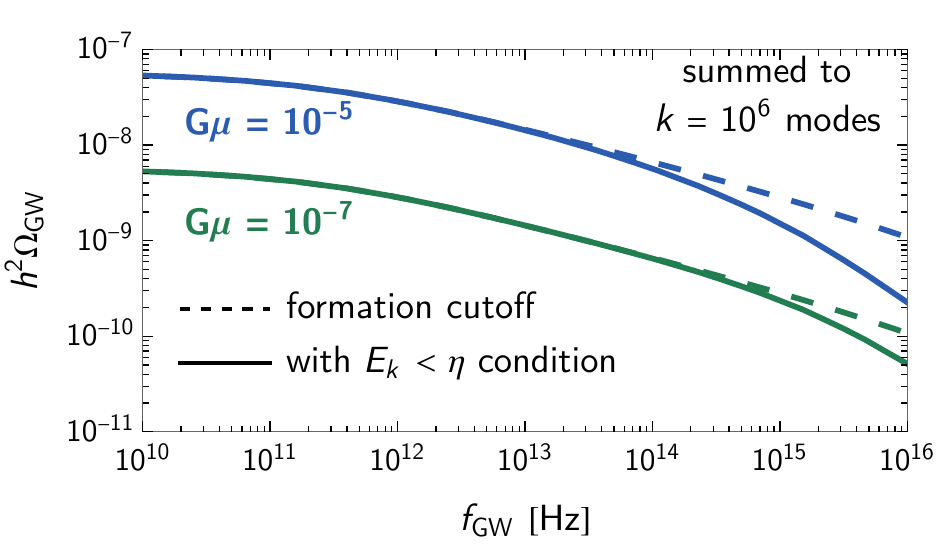}\\[-0.5em]
\caption{Top panel: The solid lines show GWB spectrum from each $k$-mode of loop oscillations, taking into account the limit on maximum emission energy; see Eq.~\eqref{eq:higher_mode_cutoff}. 
We compare this to the GWB spectra (dashed lines), assuming only the formation cutoff.
Bottom panel: the total GW spectra, summing up to the first million modes.}
\label{fig:app_kmax}
\end{figure}

A string loop of length $l$ oscillates with frequency \cite{Vilenkin:2000jqa} $\omega_k = 4 \pi k/l$ with a mode number $k$ and allows  emission of energy $E_k = \omega_k$.
The emission of energy from a loop should not change the string's state by emitting energy larger than the mass of the scalar field, i.e., $E_k < \eta$ or $l > 4\pi k/\eta$. That is, only a loop larger than $4\pi k/\eta$ supports the oscillation of mode $k$.
Using that the loop length at formation is $l \simeq \alpha t$, we obtain that a mode $k$ oscillation is allowed on a loop below the temperature 
\begin{align}
    T_k \simeq 8.3 \times 10^{14} ~ {\rm GeV} \left(\frac{G\mu}{10^{-11}}\right)^{\frac{1}{4}}\left(\frac{1}{k}\right)^{\frac{1}{2}}.
    \label{eq:higher_mode_cutoff}
\end{align}
Fig.~\ref{fig:app_kmax} compares the GWB spectra including the $E_k < \eta$ condition, i.e., we consider only loop produced after $t_i > t[\min(T_k,T_{\rm form})]$, to the simple formation cutoff, i.e., $t_i > t(T_{\rm form})$.
Fig.~\ref{fig:app_kmax}-top shows the GWB spectra from each mode $k$: $\Omega_{\rm GW}^{(k)} = k^{-4/3} \Omega_{\rm GW}^{\rm (1)}(f/k)$ with $\Omega_{\rm GW}^{\rm (1)}$ is the first-mode spectrum. For larger $k$, we see a more significant deviation between the spectra, including the effect of $T_k$ in Eq.~\eqref{eq:higher_mode_cutoff}, and those cut by the formation cutoff.
In Fig.~\ref{fig:app_kmax}-bottom, we sum the spectra from $k=1$ up to $k=10^6$. The $T_k$ effect changes the UV slope of the spectrum at frequencies higher than the formation cutoff. It makes the slope of the UV tail steeper to the asymptotic slope of $\Omega_{\rm GW}\propto f_{\rm GW}^{-1}$.
The reason is that, at higher frequencies, the contributions from large $k$-modes are more suppressed, and the first mode is responsible for the visible UV tail. 

\section{Peaked GWB}
\label{app:peak_GWB}

\begin{figure}[h]
\includegraphics[width=\linewidth]{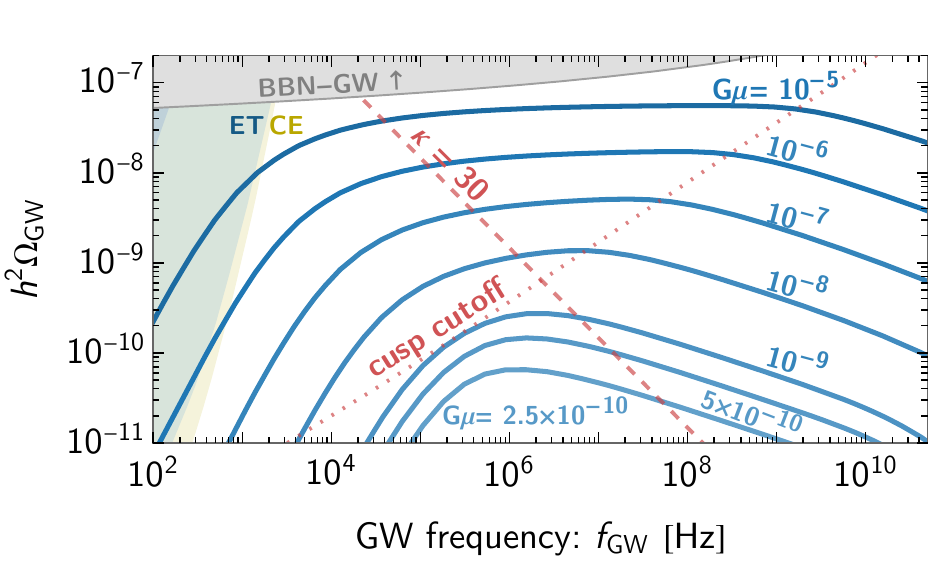}\\[-0.5em]
\caption{Local-string GWB spectra varying $G\mu$ in blue with the metastability ($\kappa = 30$) and cusp cutoffs (in red). 
The spectrum exhibits a flat plateau shape when $f_{\rm GW}^{\rm meta} < f_{\rm GW} < f_{\rm GW}^{\rm cusp}$.
For $f_{\rm GW}^{\rm meta} > f_{\rm GW}^{\rm cusp}$, the spectrum becomes a peak shape with the suppressed amplitude [see Eq.~\eqref{eq:GW_suppressed_two_cutoffs}].}
\label{fig:local_GW_GWB_combined_cutoff_app}
\end{figure}

Complementary to Fig.~\ref{fig:local_GW_GWB_combined_cutoff} which shows the combined effects of the metastability and the cusp cutoffs for a given  $G\mu$,  Fig.~\ref{fig:local_GW_GWB_combined_cutoff_app} fixes the metastability parameter $\kappa$ and varies $G\mu$ instead. We see that the GWB spectra of $G\mu$ in Eq.~\eqref{eq:peak_GW_spectrum_Gmu_sup}---corresponding to the GWB below where the two red lines cross---exhibit peak-shape feature.

\section{GWB contribution from segments}
\label{app:segments}

\begin{figure*}[t]
{\bf Metastable local-string GWB (loops + segments)}\\[-0.5em]
\includegraphics[width=0.495\linewidth]{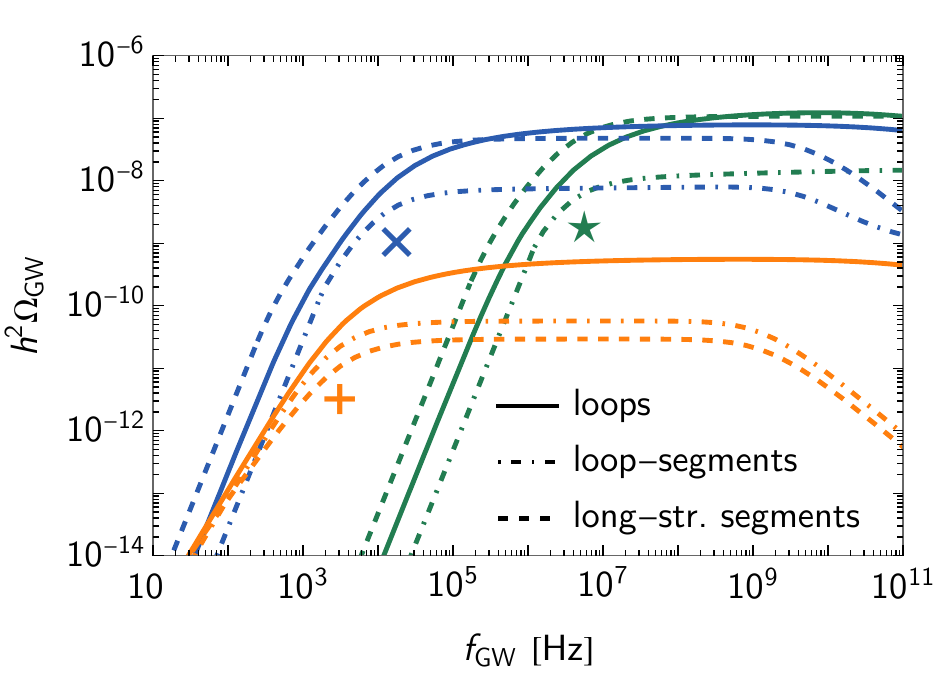}\hfill
\includegraphics[width=0.495\linewidth]{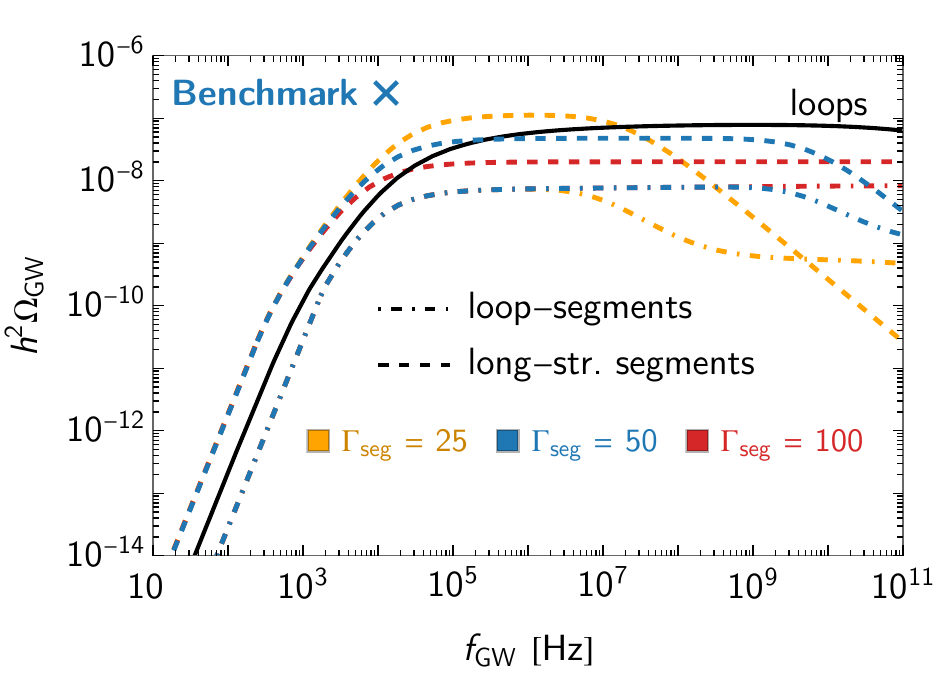}\\[-0.5em]
\caption{The GWB from metastable local strings consists of three contributions: (i) loops (solid lines) discussed in section~\ref{sec:local_metastable_network}, (ii) segments from loops (dot-dashed lines), and (iii) segments from long strings (dashed lines). 
Assuming $\Gamma_{\rm seg} = \Gamma = 50$, the left plot shows the spectra of the same benchmark scenarios as in Fig.~\eqref{fig:local_GW_GWB}, where we plot again the loops' GWB for comparison.
The right plot illustrates the effect of $\Gamma_{\rm seg}$ on the GWB from segments for the benchmark scenario ``cross".}
\label{fig:local_GW_GWB_segment}
\end{figure*}

In addition to the GWB from loops discussed in the main text, the metastable local-string segments attached to monopoles on their ends also produce GWB \cite{Martin:1996cp,Leblond:2009fq}; see the recent thorough study in \cite{Buchmuller:2021mbb}.
The string-monopole segments can be formed after time $t_{\rm brk}$ in Eq.~\eqref{eq:t_brk} via two processes: (i) segments from loops and (ii) segments from the breaking of long strings.

Similar to the GWB from loops in Eq.~\eqref{eq:GWB_spectrum_master}, the calculation for segments needs the GW emission power from a segment and the number density of segments.
The former is addressed in \cite{Martin:1996cp}, while the latter is derived from the conservation of loops' number density and the string network's energy density at time $t_{\rm brk}$ \cite{Leblond:2009fq,Buchmuller:2021mbb}.

A number density of segments produced from loops is \cite{Buchmuller:2021mbb},
\begin{align}
    n_{\rm seg}^{\rm loop} \simeq \sigma \log(\mathcal{E}^{-1}) n_{\rm loop}^{\rm brk},
    \label{eq:loop_segment_number_density}
\end{align}
where $\sigma \simeq 5$ is the numerical factor corresponding to considering many generations of segments, $\mathcal{E}$ is the suppression factor in Eq.~\eqref{eq:suppression_factor}, and $n_{\rm loop}^{\rm brk}$ is the number density of decaying loops in Eq.~\eqref{eq:local_loop_density}.

For long-string segments in radiation era, the number density reads \cite{Buchmuller:2021mbb},
\begin{align}
    n_{\rm seg}^{\rm long} \simeq \frac{\Gamma_d^2}{\xi^2} \frac{(t+t_{\rm brk})^2}{\sqrt{t^3 t_{\rm brk}}} \mathcal{E}_{\rm seg}(l,t),
    \label{eq:long_segment_number_density}
\end{align}
where $\xi$ is the long-string correlation length in unit of cosmic time (in radiation era, $\xi \approx 0.27$  \cite{Gouttenoire:2019kij}), and the suppression factor
\begin{align}
    \mathcal{E}_{\rm seg}(l,t) = e^{-\Gamma_d\left[l(t)(t+t_{\rm brk}) + \frac{1}{2}\Gamma_{\rm seg} G\mu\left(t-t_{\rm brk}\right)(t+3t_{\rm brk})\right]},
    \label{eq:suppression_factor_long_segment}
\end{align}
with $\Gamma_{\rm seg}$ being the total GW emission power from string-monopole segments which we will now discussed.

As shown in \cite{Martin:1996cp}, the GW emission power from the string-monopole segment is $\Gamma_{\rm seg}^{(k)} = 4/k$ where $k$ runs from 1 to $k_{\rm max}$. The maximum mode is $k_{\rm max} \sim \gamma_0^2$ where $\gamma_0$ is the Lorentz factor of the monopoles at the string's end. Assuming the monopoles get their kinetic energy from the string's energy, 
\begin{align}
    \gamma_0^2 \sim \frac{\mu^2 l^2}{m^2} \sim \begin{cases}
    \frac{\mu t_{\rm brk}^2}{\kappa}  \sim \frac{\exp(\pi \kappa)}{\kappa} &\textrm{(long-seg.)},\\[0.25em]
    \frac{\mu t_{i,\rm sup}^2}{\kappa} \sim \frac{\exp(\pi \kappa)}{\kappa}G\mu &\textrm{(loop-seg.)},
    \end{cases}
\end{align}
where we consider the string-monopole segment at the time $t_{\rm brk}$ (i.e., $l \sim t_{\rm brk}$ for long string's segment and $l \sim t_{i,\rm sup} \simeq 2 \Gamma G \mu t_{\rm sup}/\alpha$ for loop's segment) and use Eqs.~\eqref{eq:t_brk} and \eqref{eq:t_sup}.
The total power of GW emission from the segment is, 
\begin{align}
    \Gamma_{\rm seg} = \sum_{k=1}^{k_{\rm max}}\Gamma_{\rm seg}^{(k)} &\sim 4 \ln k_{\rm max} \label{eq:segment_GW_emission_power}\\
    &\sim \begin{cases}
    4 \pi \kappa &\textrm{(long-seg.)},\\
    4 [ \pi \kappa + \log(G\mu)] &\textrm{(loop-seg.)},
    \end{cases} \nonumber
\end{align}
where the last step approximates the discrete sum with integral. Nonetheless, many works use that the total emission power of segments is similar to the emission from a loop, $\Gamma_{\rm seg} \simeq \Gamma \simeq 50$, although we see from Eq.~\eqref{eq:segment_GW_emission_power} that $\Gamma_{\rm seg} \sim \mathcal{O}(10-10^3)$ for $\kappa \sim \mathcal{O}(1-100)$.

The GWB from loop segments is derived by replacing the loop number density and GW emission power $\Gamma^{(k)}$ in Eq.~\eqref{eq:GWB_spectrum_master} with  Eq.~\eqref{eq:loop_segment_number_density} and $\Gamma_{\rm seg}^{(k)}$, respectively. For long-string segments, we instead use Eq.~\eqref{eq:long_segment_number_density} for the number density. Note that Eq.~\eqref{eq:GWB_spectrum_master} can be used for segments as long as the GW frequency emitted by a segment is related to its length in the same way as for a loop, i.e., $f_{\rm GW}^{\rm emit} = 2 k/l$. However, as shown in \cite{Leblond:2009fq}, $f_{\rm GW}^{\rm emit}$ relates to the speed $v_0$ and acceleration $a$ of monopoles, i.e., $f_{\rm emit} = k a/(2 \gamma_0 v_0) \simeq  k (\gamma_0-1)/(\gamma_0 v_0 l)$ (assuming the length when the monopoles are at rest $l = 2(\gamma_0-1)/a$ \cite{Leblond:2009fq}). One can expect the frequency shift of the segment-GWB spectra relative to results we will now show. For accurate spectra, one might need numerical simulations of string-monopole segments to find the GW emission power, similar to what is done for loops \cite{Blanco-Pillado:2017oxo}.

Fig.~\ref{fig:local_GW_GWB_segment} shows the GWBs from metastable local strings, which decay via monopole-pair nucleation, and include all three contributions from: loops, segments from loops, and segments from long strings. See also Fig.~5 of \cite{Buchmuller:2021mbb} for other values of $G\mu$ and $\kappa$. This figure extends Fig.~\ref{fig:local_GW_GWB}, which shows only the GWB from loops. For simplicity (as in \cite{Buchmuller:2021mbb}), the left plot fixes $\Gamma_{\rm seg} = \Gamma = 50$ for both types of segments; that is, $k_{\rm max} \simeq 2.7 \times 10^5$.
The long-string segment's contribution becomes large for large $G\mu$ and gives a significant correction to the GWB, while the loop segment's contribution remains smaller than that of loops. If the IR tail, the metastability cutoff, or the flat plateau is observed, this correction from segments can introduce the uncertainty in reconstructing parameters $\{G\mu, \kappa\}$.

We observe that, for $G\mu \gtrsim 10^{-9}$, the long-string segment's contribution dominates the IR tail; see benchmark ``$+$". For $G\mu \simeq 10^{-5}$, it is $\sim 10$ times larger than the IR tail of loop contribution by $\mathcal{O}(10)$.
Using Eq.~\eqref{eq:GWB_stable}, this leads to the uncertainty in $G\mu$ of order  $\mathcal{O}(10^2)$ or equivalently in energy scale $\eta$ of order $\mathcal{O}(10)$. Moreover, the flat plateau is dominated by the long-string segments when $G\mu \gtrsim 6 \times 10^{-5}$; see benchmark ``$\star$".
We also see that the metastability cut-off moves to lower frequency by $\sim \mathcal{O}(10-10^2)$ for the largest $G\mu \sim 10^{-5}-10^{-4}$ considered in this work. Identifying the cutoff using Eq.~\eqref{eq:metastable_cutoff_freq}, the segment contribution could lead to misidentifying $\kappa$ by a factor of $\sim3-6$ smaller.

Finally, we emphasize that the segments' contributions should be taken with care when confronting real data analysis. As shown on the right panel of Fig.~\ref{fig:local_GW_GWB_segment}, the amplitude of the long-string segments' GWB depends on $\Gamma_{\rm seg}$ that is subjected to theoretical uncertainties. Moreover, the flat plateau of the long-string segments\footnote{Unlike loops' GWB, this flat spectrum does not arise from the fact that this is a long-lasting source of GW. The scale-invariant spectrum comes from the summation of modes and from the fact that the amplitude for each mode scales as $k^{-1}$.} extends up to the UV cutoff, which depends on $k_{\rm max}$ (or $\Gamma_{\rm seg}$).

\section{Modified causality tail of heavy-axion string GWB}
\label{app:axion_MD_IR_tail}

\begin{figure}[h!]
\includegraphics[width=\linewidth]{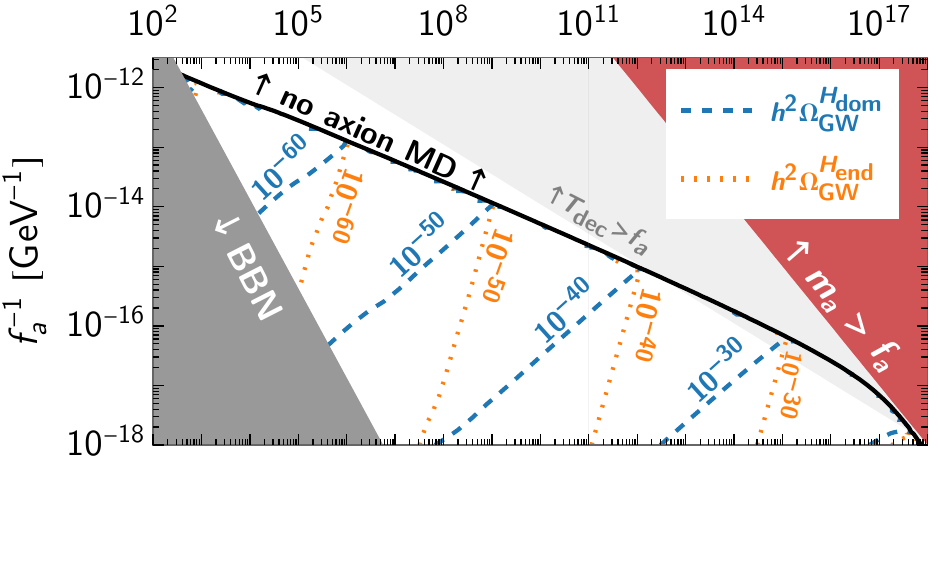}\\[-4em]
\includegraphics[width=\linewidth]{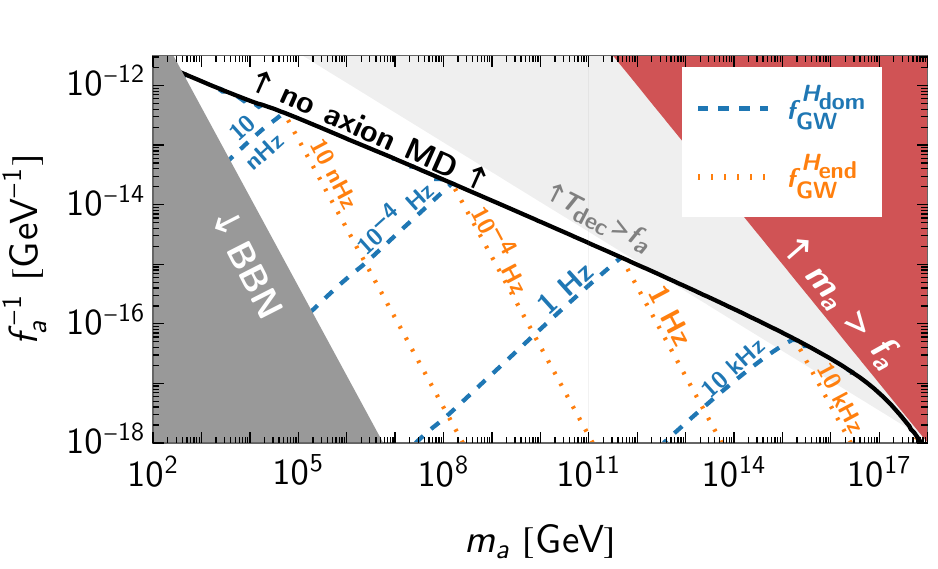}\\[-0.5em]
\caption{In the axion parameter space, the contour lines show amplitudes and frequencies of the modified causality tail of the axion-string GWB due to the axion MD era.}
\label{fig:axion_MD_IR_tail}
\end{figure}

Another effect of the axion MD era on the GWB from global strings is to modify the causality tail from the $\Omega_{\rm GW} \propto f_{\rm GW}^3$ scaling which assumes the radiation-domination era into the $\Omega_{\rm GW} \propto f_{\rm GW}$ during the matter era \cite{Hook:2020phx, Racco:2022bwj, Franciolini:2023wjm}.
The modification happens at the frequencies corresponding to the horizon scale during the axion MD era, i.e., $f_{\rm GW}^H = H [a(H)/a_0]$ with $H_{\rm dom}> H > H_{\rm end}$.
We have
\begin{align}
    f_{\rm GW}^{H_{\rm end}} \simeq 18 ~ {\rm nHz} \left[\frac{g_*(T_{\rm end})}{106.75}\right]^\frac{1}{2} \left[\frac{106.75}{g_{*s}(T_{\rm end})}\right]^\frac{1}{3}\left[\frac{T_{\rm end}}{1 ~ {\rm GeV}}\right],
    \label{eq:GW_freq_causality_Tend}
\end{align}
and
\begin{align}
  f_{\rm GW}^{H_{\rm dom}} \simeq f_{\rm GW}^{H_{\rm end}} \, \mathcal{B}^{-\frac{1}{2}}.
    \label{eq:GW_freq_causality_Tend2}
\end{align}
Fig.~\ref{fig:axion_MD_IR_tail} shows the GW amplitudes of the modified causality tail at frequencies $f_{\rm GW}^{H_{\rm end}}$ and $f_{\rm GW}^{H_{\rm dom}}$. Although they fall in the sub-kHz range, the amplitude is too small to be detected by any current or future GW experiments.

\newpage
\FloatBarrier


\bibliographystyle{JHEP}
\bibliography{uhf.bib}

\end{document}